\documentclass[cameraready]{Interspeech}
\title{Dr. SHAP-AV: Decoding Relative Modality Contributions via Shapley Attribution in Audio-Visual Speech Recognition}

\author[affiliation={\heartsuit}]{Umberto}{Cappellazzo}
\author[affiliation={\heartsuit,\spadesuit}]{Stavros}{Petridis}
\author[affiliation={\heartsuit,\spadesuit}]{Maja}{Pantic}
\address{
    $^\heartsuit$ Imperial College London, UK \\
    $^\spadesuit$ NatWest AI Research, UK
}

\email{u.cappellazzo@imperial.ac.uk}


\usepackage{comment}
\definecolor{darkmaroon}{RGB}{100, 16, 26}
\definecolor{teagreen}{rgb}{0.82, 0.94, 0.75}
\definecolor{pastelviolet}{RGB}{234, 229, 246}
\usepackage{soul}
\definecolor{tealclean}{HTML}{029386}
\definecolor{coralnoisy}{HTML}{FF7F50}
\sethlcolor{tealclean!30}
\newcommand{\hlclean}[1]{\sethlcolor{tealclean!30}\hl{#1}}
\newcommand{\hlnoisy}[1]{\sethlcolor{coralnoisy!30}\hl{#1}}
\usepackage{tikz}
\newcommand{\circmark}[1]{\tikz[baseline=(char.base)]{\node[shape=circle, draw=darkmaroon, fill=darkmaroon, text=white, inner sep=1.5pt, font=\bfseries\small] (char) {#1};}}

\usepackage[utf8]{inputenc} % allow utf-8 input
\usepackage[T1]{fontenc}    % use 8-bit T1 fonts
\usepackage{hyperref}       % hyperlinks
\usepackage{url}            % simple URL typesetting
\usepackage{booktabs}       % professional-quality tables
\usepackage{amsfonts}       % blackboard math symbols
\usepackage{nicefrac}       % compact symbols for 1/2, etc.
\usepackage{microtype}      % microtypography
\usepackage{xcolor}         % colors
\usepackage{times}
\usepackage{latexsym}
\usepackage{tcolorbox}
\tcbuselibrary{skins}

\usepackage{amsmath}
\usepackage{cite}
\usepackage{colortbl}
\usepackage{adjustbox}
\usepackage{arydshln}
\usepackage{multirow}
\usepackage{pifont}% http://ctan.org/pkg/pifont

\usepackage{amssymb}
\usepackage{algorithm}
\usepackage{listings}
\usepackage{placeins}
\usepackage[accsupp]{axessibility}
\usepackage{nccmath}

\usepackage{caption} 
\usepackage{subcaption}
\usepackage{makecell}
\usepackage{wrapfig}

\usepackage{mathtools}

\usepackage{inconsolata}

\begin{document}

\maketitle

\begin{abstract}
Audio-Visual Speech Recognition (AVSR) leverages both acoustic and visual information for robust recognition under noise. However, how models balance these modalities remains unclear. We present Dr.\ SHAP-AV, a framework using Shapley values to analyze modality contributions in AVSR. Through experiments on six models across two benchmarks and varying SNR levels, we introduce three analyses: \texttt{Global} \texttt{SHAP} for overall modality balance, \texttt{Generative} \texttt{SHAP} for contribution dynamics during decoding, and \texttt{Temporal} \texttt{Alignment} \texttt{SHAP} for input-output correspondence. Our findings reveal that models shift toward visual reliance under noise yet maintain high audio contributions even under severe degradation. Modality balance evolves during generation, temporal alignment holds under noise, and SNR is the dominant factor driving modality weighting. These findings expose a persistent audio bias, motivating ad-hoc modality-weighting mechanisms and Shapley-based attribution as a standard AVSR diagnostic. \textbf{Project website}: \url{https://umbertocappellazzo.github.io/Dr-SHAP-AV}

\end{abstract}

\section{Introduction}
\label{sec:introduction}

Speech recognition systems are widely deployed across a range of real-world applications, yet they remain vulnerable to acoustic noise, where background noise and competing speech can severely degrade recognition accuracy~\cite{barker2018fifth, prabhavalkar2023end, radford2023robust}. Audio-Visual Speech Recognition (AVSR)~\cite{noda2015audio, afouras2018deep, petridis2018end, autoavsr} addresses this limitation by complementing acoustic signals with visual cues, most notably lip movements, which are inherently immune to acoustic corruption. By exploiting the natural correlation between speech and the articulatory motion of the speaker's lips, AVSR systems maintain robust recognition even when audio quality degrades.

The development of AVSR has proceeded through several distinct research directions. Early approaches relied on modality-specific encoders and handcrafted fusion strategies~\cite{noda2015audio, meutzner2017improving, mroueh2015deep}, while the introduction of Transformer-based architectures~\cite{vaswani2017attention} brought substantial performance improvements~\cite{afouras2018deep, petridis2018audio}. A dominant line of research has since centered on self-supervised learning~\cite{gui2024survey}, leveraging large-scale unlabeled audio-visual data to pre-train encoders that capture the intrinsic correlation between the two modalities, even with limited labeled data~\cite{shi2022avhubert, shi2022avhubertnoise, haliassos2022jointly, haliassos2024unified}. Other paradigms have also emerged, such as ASR-to-AVSR distillation~\cite{autoavsr, rouditchenko2024whisper}, which uses pre-trained auditory speech recognition (ASR) models to automatically annotate large-scale audio-visual datasets, and cross-modal complementarity approaches~\cite{hong2022visual, hong2023watch} that explicitly model the synergistic relationship between auditory and visual streams.  More recently, the rise of Large Language Models (LLMs)  and their multimodal extensions~\cite{shi2024eagle, yue2024pangea, bai2025qwen3, liu2025thinksound, li2025watch} has opened a new frontier: several works have successfully aligned speech representations with LLMs for AVSR as well as the unimodal tasks ASR and visual speech recognition (VSR), attaining state-of-the-art recognition performance~\cite{llamaavsr, yeo2025mms, cappellazzo2025scaling, omniavsr, cappellazzo2025mitigating, zhang2025adapting, yeo2024visual, fathullah2024prompting}.

\begin{figure}[t]
    \centering
    \includegraphics[width=\linewidth]{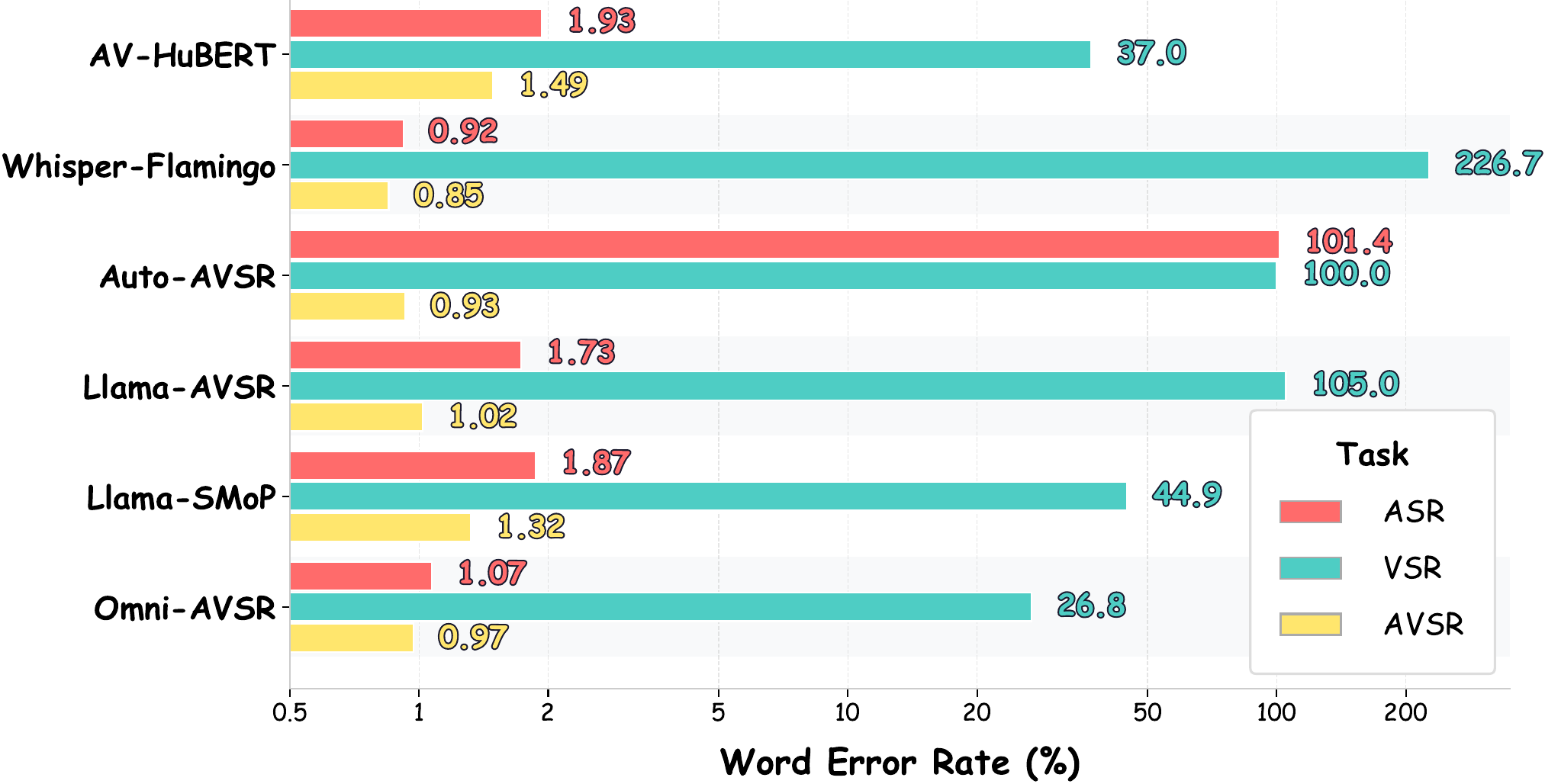}
    \caption{WER comparison ($\downarrow$) when ablating audio or video modalities across six state-of-the-art AVSR models in clean conditions. Note that the x-axis is displayed on a logarithmic scale.}
    \label{fig:drop_tokens}
    \vspace{-0.5cm}
\end{figure}

While integrating visual lip information is clearly beneficial in noisy acoustic conditions, its contribution in clean conditions is far less evident. This can be attributed to the inherent asymmetry between the two modalities: audio directly encodes the full speech signal at high temporal resolution, whereas visual lip movements are captured at lower frame rates and carry intrinsically more ambiguous information, as many phonemes share near-identical visual appearance. As a consequence, during joint training, models tend to latch onto the audio modality, which is easier and faster to learn, while the visual encoder receives comparatively weak gradient updates~\cite{peng2022balanced, zhang2024multimodal, dai2024study}. A telling symptom of this bias is observable in the performance reported by state-of-the-art AVSR models: when evaluated under clean conditions, the gap between ASR and AVSR systems is consistently marginal~\cite{autoavsr, llamaavsr, rouditchenko2024whisper, haliassos2024unified}.     

To directly quantify this modality reliance, we analyze six recent AVSR models by independently dropping all audio or video tokens at inference time, effectively reducing the model to either an ASR-only or VSR-only system, and compare against the full AVSR performance. The results are illustrated in Figure~\ref{fig:drop_tokens}. Across all models, removing video tokens while retaining audio (ASR condition) results in WER values that remain close to the full AVSR system, confirming that audio alone largely drives recognition in clean conditions. In contrast, removing audio tokens while retaining video (VSR condition) causes a dramatic degradation across the board, with WER increasing by one to two orders of magnitude for most models. This asymmetry is particularly stark for the four models trained exclusively on the AVSR task (Llama-AVSR~\cite{llamaavsr}, Llama-SMoP~\cite{cappellazzo2025scaling}, Whisper-Flamingo~\cite{rouditchenko2024whisper}, and Auto-AVSR~\cite{autoavsr}), which, having never been explicitly trained to rely on visual input alone, collapse almost entirely when audio is absent. Interestingly, AV-HuBERT~\cite{shi2022avhubert} and Omni-AVSR~\cite{omniavsr}, both trained with a multi-task objective that includes the VSR task, exhibit comparatively stronger visual-only performance, suggesting that explicit VSR supervision is a key factor in encouraging models to extract meaningful information from the visual stream. 

These observations raise a fundamental \textbf{question}:
\begin{tcolorbox}[enhanced, drop fuzzy shadow, colback=darkmaroon!15, colframe=darkmaroon, title= Research Question:, fonttitle=\bfseries] How do AVSR models balance audio and visual contributions, and what factors shape this balance across acoustic conditions, the decoding process, and input characteristics? \end{tcolorbox}
%How much does each modality truly contribute to AVSR, and does this balance shift across different models, noise levels, and input conditions?
\textit{Developing a principled understanding of modality contribution in AVSR models is therefore the central goal of this work}.

In the literature, little effort has been devoted to studying modality contributions in AVSR, as most works have focused on reducing WER to achieve state-of-the-art performance. For example, Dai et al.~\cite{dai2024study} investigate why AVSR systems are sensitive to missing video frames, tracing the root cause to an excessive modality bias toward audio induced by dropout. Lin et al.~\cite{lin2025uncovering} examine how AVSR systems exploit visual information through the lens of human speech perception, revealing that current methods do not fully leverage visual cues. Papadopoulos et al.~\cite{papadopoulos2025interpreting} study how the visual modality contributes to AVSR by applying interpretability techniques to examine how visemes are encoded in AV-HuBERT. However, these works focus on individual models or provide empirical observations without formal mathematical frameworks, and none encompasses a comprehensive analysis across multiple state-of-the-art AVSR architectures.

To address these limitations, we turn to \textit{Shapley values} from cooperative game theory~\cite{shapley1953value}, which provide a rigorous mathematical framework for quantifying modality contributions. Unlike heuristic analyses, Shapley values offer theoretically grounded attributions satisfying desirable properties such as efficiency, symmetry, and fair credit assignment~\cite{lundberg2017unified}. Crucially, they are \textit{performance-agnostic}: they quantify how models utilize inputs based solely on prediction behavior, not on whether the output is correct. This makes them ideally suited for analyzing AVSR, where we aim to understand modality utilization regardless of transcription accuracy. 

Shapley-based attribution has recently gained traction in the multimodal learning community. Parcalebescu et al.~\cite{parcalabescu2023mm} introduced MM-SHAP to measure how vision-language models utilize image and text modalities in encoder-based classification tasks. This framework was subsequently extended to autoregressive architectures: Parcalebescu et al.~\cite{parcalabescu2025decoders} applied Shapley-based attribution to vision-language decoders, while Morais et al.~\cite{morais2025investigating} investigated modality contributions in audio-based LLMs for music understanding. Building on these recent advances, we leverage Shapley values to systematically analyze audio-visual modality contributions in speech recognition.

We present \textbf{Dr.\ SHAP-AV}, a comprehensive framework for \textbf{D}ecoding \textbf{R}elative modality contributions via \textbf{SHAP}ley attribution in \textbf{A}udio-\textbf{V}isual speech recognition. Our work extends prior Shapley-based approaches in several key directions:

\begin{itemize}
    \item \textbf{Analysis Across Acoustic Conditions.} Prior Shapley-based studies analyze modality contributions under fixed input conditions. In contrast, AVSR offers a natural testbed for examining how models adjust their audio-visual balance when input quality varies. We systematically analyze contributions across SNR levels from clean speech to $-10$~dB, revealing how models adapt under acoustic degradation.
    \item \textbf{Extension to Cross-Attention-based Architectures.} Prior Shapley-based analyses have been limited to \textit{LLM}-based architectures~\cite{parcalabescu2025decoders, morais2025investigating}. We extend this framework also to \textit{cross-attention-based} encoder-decoder models, which represent one of the dominant paradigms in AVSR.
    \item \textbf{Multi-Granularity Analysis.} Beyond global modality contributions~\cite{parcalabescu2023mm}, we introduce two new metrics: \texttt{Generative} \texttt{SHAP} to track how modality reliance evolves during token generation, and \texttt{Temporal} \texttt{Alignment} \texttt{SHAP} to examine the correspondence between input feature positions and output tokens. We also analyze how utterance duration, different acoustic noise types, and input difficulty affect modality weighting.
\end{itemize}

Dr.\ SHAP-AV provides a unified analysis of modality behavior in AVSR along \textbf{three complementary axes}: \textbf{(i)} modality weighting under controlled SNR shifts, \textbf{(ii)} generation-time contribution dynamics during autoregressive decoding, and \textbf{(iii)} temporal alignment between input modalities and output tokens. Together, these perspectives move beyond static attribution and offer a structured characterization of cross-modal behavior in modern AVSR systems. We conduct extensive experiments on six state-of-the-art models spanning encoder-decoder (AV-HuBERT~\cite{shi2022avhubert}, Auto-AVSR~\cite{autoavsr}, Whisper-Flamingo~\cite{rouditchenko2024whisper}) and LLM-based (Llama-AVSR~\cite{llamaavsr}, Llama-SMoP~\cite{cappellazzo2025scaling}, Omni-AVSR~\cite{omniavsr}) architectures on the two main AVSR benchmarks LRS2 and LRS3. Our key findings are:

\begin{tcolorbox}[enhanced, drop fuzzy shadow, colback=darkmaroon!15, colframe=darkmaroon, title=Summary of Findings:, fonttitle=\bfseries]
\begin{enumerate}[leftmargin=*, labelsep=0.5em, itemsep=0.3em]
    \item[\circmark{1}] AVSR models dynamically shift toward visual reliance as audio quality deteriorates, yet maintain surprisingly high audio contributions ($38$-$46$\%) even at $-10$~dB SNR, where substantially stronger visual dominance would be expected.
    
    \item[\circmark{2}] Modality contribution evolves during generation: Whisper-Flamingo and Omni-AVSR progressively increase audio reliance as decoding proceeds, while AV-HuBERT maintains stable balance throughout.
    
    \item[\circmark{3}] Both modalities maintain temporal alignment between input features and output tokens, and this structure remains robust under acoustic degradation.
    
    \item[\circmark{4}] Different noise types induce varying degrees of visual reliance, with more challenging conditions producing larger shifts toward the visual modality.

    \item[\circmark{5}] Utterance duration affects modality contributions differently across architectures, revealing no consistent pattern but rather model-specific trends shaped by architectural design.

    \item[\circmark{6}] Acoustic conditions are the dominant factor driving modality balance, while recognition difficulty has minimal effect within each SNR level.

\end{enumerate}
\end{tcolorbox}

Overall, our analysis provides actionable insights into how AVSR models integrate multimodal information, revealing both adaptive capabilities and persistent biases that motivate the development of explicit modality-weighting mechanisms and encouraging future works to adopt Shapley-based attribution as a standard diagnostic for AVSR.

\section{Methodology}
\label{sec:method}

In this section, we describe \textbf{Dr.\ SHAP-AV}, our framework for \textbf{D}ecoding \textbf{R}elative modality contributions via \textbf{SHAP}ley attribution in \textbf{A}udio-\textbf{V}isual speech recognition. It builds on Shapley values from cooperative game theory~\cite{shapley1953value}, which provide a principled method for fairly attributing model predictions to input features. We organize this section as follows: \textbf{1)} we introduce the necessary background on Shapley values, then \textbf{2)} describe how we compute and adapt them to autoregressive AVSR models during inference, and finally \textbf{3)} present three complementary metrics to analyze modality contributions at different granularities.

\begin{figure*}[t]
    \centering
    \includegraphics[width=0.8\textwidth]{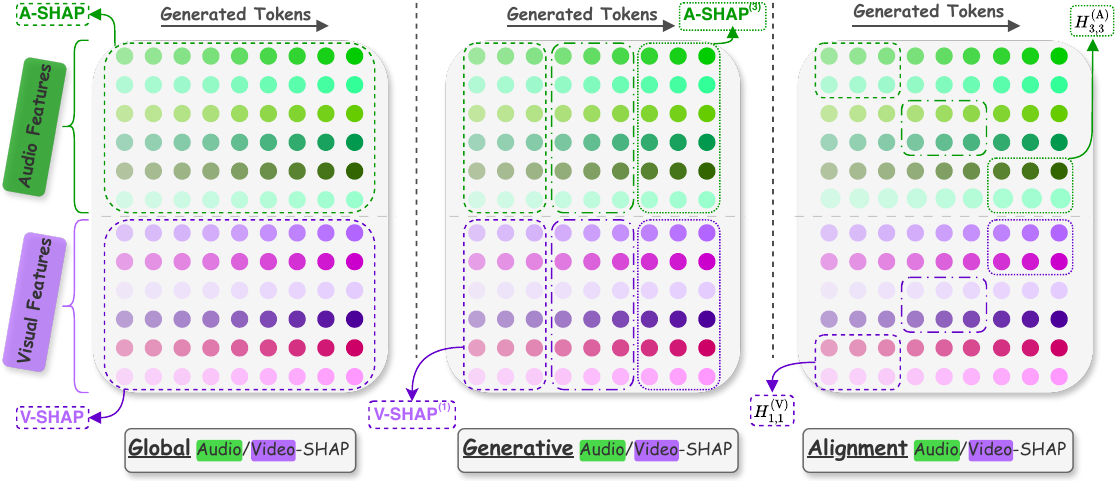}
    \caption{Overview of the three proposed SHAP-based analyses in Dr.\ SHAP-AV. From the Shapley matrix $\bm{\Phi}$, which captures the contribution of each input feature (rows) to each generated token (columns), we compute: \textbf{(Left)} \texttt{Global} \texttt{SHAP}, which aggregates contributions across all features and tokens to quantify overall modality balance; \textbf{(Middle)} \texttt{Generative} \texttt{SHAP}, which tracks modality contribution dynamics across token generation stages; and \textbf{(Right)} \texttt{Temporal} \texttt{Alignment} \texttt{SHAP}, which examines the correspondence between input feature positions and output token positions.}
    \label{fig:main_diagram}
    \vspace{-0.5cm}
\end{figure*}

\subsection{Background: Shapley Values for Feature Attribution}
Shapley values originate from cooperative game theory, providing a principled method for fairly distributing a total payoff among contributing players~\cite{shapley1953value}. In the context of machine learning, Shapley values have been widely adopted for feature attribution, where the ``players'' are input features or tokens and the ``payoff'' is the model's prediction ~\cite{vstrumbelj2014explaining, lundberg2017unified, covert2021explaining, beechey2023explaining}.

Given a model $f: \mathcal{X} \rightarrow \mathbb{R}$ and an input $\bm{x} = (x_1, x_2, \ldots, x_n) \in \mathcal{X}$ defined over a set of $n$ features (or tokens) $\mathcal{F} = \{1, \ldots, n\}$, the \emph{characteristic function} $f_{\bm{x}}(\mathcal{C})$ represents the model's expected prediction when only the features in the subset $\mathcal{C} \subseteq \mathcal{F}$ are known~\cite{vstrumbelj2014explaining}:
\begin{equation}
    f_{\bm{x}}(\mathcal{C}) = \mathbb{E}\left[ f(\bm{X}) \mid \bm{X}_{\mathcal{C}} = \bm{x}_{\mathcal{C}} \right],
    \label{eq:characteristic}
\end{equation}
where $\bm{x}_{\mathcal{C}}$ denotes the subset of values $\{x_i : i \in \mathcal{C}\}$. The characteristic function captures how the model's prediction changes as different subsets of features are available. Features not being part of the subset are masked (i.e., set to $0$). In particular, the difference $f_{\bm{x}}(\mathcal{F}) - f_{\bm{x}}(\emptyset)$ quantifies the total change in prediction when all features are known versus when none are known. 

The \emph{Shapley value} $\phi_i(f_{\bm{x}})$ quantifies the contribution of feature $i$ to the prediction $f(\bm{x})$, and can
be \textbf{positive} (increases the model prediction) or \textbf{negative} (decreases it) or \textbf{zero} (no effect). It is defined as the weighted average of marginal contributions across all possible coalitions~\cite{shapley1953value}:
\begin{equation}
    \phi_i(f_{\bm{x}}) = \sum_{\mathcal{C} \subseteq \mathcal{F} \setminus \{i\}} \frac{|\mathcal{C}|! \cdot (|\mathcal{F}| - |\mathcal{C}| - 1)!}{|\mathcal{F}|!} \left[ f_{\bm{x}}(\mathcal{C} \cup \{i\}) - f_{\bm{x}}(\mathcal{C}) \right].
    \label{eq:shapley}
\end{equation}

The Shapley value is the unique attribution method satisfying four desirable axioms: \emph{efficiency} (contributions sum to the total payoff), \emph{symmetry} (identical features receive identical attributions), \emph{linearity} (additivity across games), and \emph{null player} (irrelevant features receive zero attribution)~\cite{shapley1953value, lundberg2017unified}. Crucially, Shapley values are \textbf{performance-agnostic}: they measure contribution to the model’s internal predictive distribution, not to correctness relative to ground truth. This makes them ideally suited for analyzing AVSR models, where we aim to understand how audio and visual modalities each contribute to token generation, regardless of whether their fusion ultimately yields an accurate transcription.

Computing exact Shapley values requires evaluating $2^{|\mathcal{F}|}$ coalitions, which becomes intractable for high-dimensional inputs. Following~\cite{vstrumbelj2014explaining} and~\cite{lundberg2017unified}, we approximate Shapley values using \emph{Permutation SHAP}, which samples random permutations $\pi \in \Pi(\mathcal{F})$ and estimates each feature's marginal contribution:
\begin{equation}
    \phi_i(f_{\bm{x}}) \approx \frac{1}{M} \sum_{m=1}^{M} \left[ f_{\bm{x}}(\mathcal{P}_i^{\pi_m} \cup \{i\}) - f_{\bm{x}}(\mathcal{P}_i^{\pi_m}) \right],
    \label{eq:permutation_shap}
\end{equation}
where $\Pi(\mathcal{F})$ is the set of all permutations of $\mathcal{F}$, $\mathcal{P}_i^{\pi}$ denotes the set of features preceding $i$ in permutation $\pi$, and $M$ is the number of sampled permutations. We also report results with \emph{Sampling SHAP}~\cite{strumbelj2010efficient, castro2009polynomial}, which approximates the Shapley value sum via Monte Carlo sampling of feature coalitions. Both methods provide unbiased estimates that converge to the true Shapley values. We set the number of sampled coalitions to $M = 2000$ to ensure stable and fair attribution across modalities~\cite{parcalabescu2023mm, parcalabescu2025decoders, morais2025investigating}. Both methods are implemented using the \texttt{shap} library~\cite{lundberg2017unified}.

%Despite their theoretical appeal, exact Shapley value computation is intractable, scaling as $\mathcal{O}(2^{|\mathcal{F}|})$ with the number of features. For AVSR models with hundreds of audio and visual feature tokens, we must therefore rely on approximation methods. We experiment with two sampling-based algorithms from the SHAP library~\cite{lundberg2017unified}: the \texttt{PermutationExplainer}~\cite{vstrumbelj2014explaining}, which estimates Shapley values by averaging marginal contributions over random feature orderings, and the \texttt{SamplingExplainer}~\cite{castro2009polynomial}, which approximates the Shapley value sum via Monte Carlo sampling of feature coalitions. Both methods provide unbiased estimates that converge to the true Shapley values. We set the number of sampled coalitions to $M = 2000$ to ensure stable and fair attribution across modalities.

\subsection{Shapley Values for Audio-Visual Speech Recognition}

While~\cite{parcalabescu2023mm} introduced MM-SHAP for vision-language encoders,~\cite{parcalabescu2025decoders} recently extended this framework to autoregressive vision-language decoders by computing Shapley values for each generated token. We build upon their approach and adapt it to the audio-visual speech recognition domain. Notably, our work extends beyond LLM-based architectures. While recent AVSR models concatenate audio, visual, and text tokens into a single sequence fed to a large language model~\cite{llamaavsr, yeo2025mms, omniavsr, zhang2025adapting, cappellazzo2025mome}, the dominant paradigm in AVSR employs \emph{encoder-decoder} architectures ~\cite{shi2022avhubert, shi2022avhubertnoise, autoavsr, rouditchenko2024whisper, haliassos2024unified, kim2025mohave}, where the audio and visual features are cross-attended by the decoder. We therefore extend the Shapley-based attribution framework to accommodate both architectural families, enabling a unified analysis of modality contributions across the AVSR landscape.

Consider a pre-trained AVSR model $f$ that generates a sequence of $T$ output tokens $\bm{y} = (y_1, \ldots, y_T)$ given multimodal input features during inference. The input consists of:  \textbf{audio features} $\bm{x}^A = (x_1^A, \ldots, x_{N_A}^A)$, extracted from the speech signal, and \textbf{visual features} $\bm{x}^V = (x_1^V, \ldots, x_{N_V}^V)$, extracted from lip video frames. We note that \textit{prompt} and \textit{special tokens} are \textbf{not} taken into account for our analysis.

Let $\mathcal{F} = \mathcal{A} \cup \mathcal{V}$ denote the complete set of input features, where $\mathcal{A} = \{1, \ldots, N_A\}$ indexes audio features, $\mathcal{V} = \{N_A + 1, \ldots, N\}$ indexes visual features, and $N = N_A + N_V$ is the total number of features. For LLM-based models, these features correspond to audio and visual tokens after being projected into the LLM space; for encoder-decoder models, they correspond to the representations from the respective modality encoders.

For each generated token $y_t$, we define the characteristic function $f_{\bm{x}}^t: 2^{\mathcal{F}} \rightarrow \mathbb{R}$ as the expected log-probability of generating $y_t$ when only features in $\mathcal{C} \subseteq \mathcal{F}$ are observed:
\begin{equation}
    f_{\bm{x}}^t(\mathcal{C}) = \mathbb{E}\left[ \log p(y_t \mid \bm{X}_{\mathcal{C}}, y_{<t}) \mid \bm{X}_{\mathcal{C}} = \bm{x}_{\mathcal{C}} \right],
    \label{eq:token_characteristic}
\end{equation}
where $\bm{x}_{\mathcal{C}}$ denotes features with indices in $\mathcal{C}$, and features in $\mathcal{F} \setminus \mathcal{C}$ are masked. In practice, following~\cite{vstrumbelj2014explaining, lundberg2017unified, parcalabescu2023mm, parcalabescu2025decoders}, we approximate this expectation by masking features and performing a single forward pass, rather than explicitly marginalizing over absent features. The Shapley value of feature $i$ for output token $t$ is~\cite{parcalabescu2025decoders}:
\begin{equation}
    \phi_{i,t} = \sum_{\mathcal{C} \subseteq \mathcal{F} \setminus \{i\}} \frac{|\mathcal{C}|! \cdot (N - |\mathcal{C}| - 1)!}{N!} \left[ f_{\bm{x}}^t(\mathcal{C} \cup \{i\}) - f_{\bm{x}}^t(\mathcal{C}) \right].
    \label{eq:token_shapley}
\end{equation}

Computing Eq.~\eqref{eq:token_shapley} for all features and tokens yields the \textbf{Shapley matrix} $\bm{\Phi} \in \mathbb{R}^{N \times T}$ with entries $\phi_{i,t}$, capturing the contribution of each input feature $i$ to each generated token $t$. This matrix serves as the foundation for all subsequent analyses.

\subsection{Dr.\ SHAP-AV: Three Levels of Analysis}
We propose three complementary analyses that operate on the Shapley matrix $\bm{\Phi}$ at different levels of granularity, each addressing a distinct research question about modality contributions in AVSR (see Figure~\ref{fig:main_diagram} for a visual overview).

\subsubsection{\texttt{Global} \texttt{SHAP}}
Following~\cite{parcalabescu2023mm, parcalabescu2025decoders}, we aggregate Shapley values across all features and tokens to quantify overall modality contributions. The global audio and visual contributions are defined as:
\begin{align}
    \text{A-SHAP} = \frac{\displaystyle\sum_{j \in \mathcal{A}} \sum_{t=1}^{T} |\phi_{j,t}|}{\displaystyle\sum_{j \in \mathcal{F}} \sum_{t=1}^{T} |\phi_{j,t}|} \quad 
    \text{V-SHAP} = \frac{\displaystyle\sum_{j \in \mathcal{V}} \sum_{t=1}^{T} |\phi_{j,t}|}{\displaystyle\sum_{j \in \mathcal{F}} \sum_{t=1}^{T} |\phi_{j,t}|}.
\end{align}

We use absolute values $|\phi_{j,t}|$ following~\cite{parcalabescu2023mm}, as we are interested in the \emph{magnitude} of each feature's contribution regardless of whether it increases or decreases the prediction probability. It holds that A-SHAP $= 1$ $-$ V-SHAP. A-SHAP $= 0.5$ indicates balanced modality usage; A-SHAP $> 0.5$ indicates audio dominance; A-SHAP $< 0.5$ indicates visual dominance.

\subsubsection{\texttt{Generative} \texttt{SHAP}}
\label{subsec:generative_shap}
While A/V-SHAP metrics provide a useful summary of overall modality contributions, they collapses all information into a single scalar value, potentially obscuring important dynamics in how models utilize different modalities throughout the generation process. We therefore extend them to capture how modality contributions evolve across autoregressive decoding. A natural question arises: \emph{Does the model rely on audio and visual information uniformly across all generation stages, or does the balance shift as decoding progresses?}

To answer this, we introduce \texttt{Generative} \texttt{SHAP}, which extends \texttt{Global} \texttt{SHAP} by tracking modality contributions across the generation process. Rather than analyzing individual tokens, which can be noisy and difficult to interpret, we group consecutive tokens into temporal windows. Specifically, we divide the generated sequence into $W$ (e.g, $W = 3$ in Figure~\ref{fig:main_diagram}) windows $\{\mathcal{T}_1, \ldots, \mathcal{T}_W\}$, where each window $\mathcal{T}_w$ contains tokens corresponding to a fixed percentage of the output sequence. For each window, we compute the modality contributions:
\begin{align}
    \text{A-SHAP}^{(w)} &= \frac{\displaystyle\sum_{j \in \mathcal{A}} \sum_{t \in \mathcal{T}_w} |\phi_{j,t}|}{\displaystyle\sum_{j \in \mathcal{F}} \sum_{t \in \mathcal{T}_w} |\phi_{j,t}|}, \label{eq:gen_ashap} \\[2mm]
    \text{V-SHAP}^{(w)} &= 1 - \text{A-SHAP}^{(w)}. \label{eq:gen_vshap}
\end{align}

Aggregating across all windows yields audio $(\text{A-SHAP}^{(1)}, \ldots, \text{A-SHAP}^{(W)})$ and video $(\text{V-SHAP}^{(1)}, \ldots, \text{V-SHAP}^{(W)})$ trajectories that reveal \textbf{dynamic modality reliance} during token generation. These trajectories capture whether models rely more heavily on audio when generating initial versus final tokens, or whether visual features become more important at certain stages, reflecting how the temporal structure of speech maps onto the generation process.

\subsubsection{\texttt{Temporal} \texttt{Alignment} \texttt{SHAP}}
\label{sec:alignment_shap}

Finally, we propose \texttt{Temporal} \texttt{Alignment} \texttt{SHAP} to investigate whether temporal correspondence exists between input feature positions and output token positions. Specifically, we ask: \emph{Do early input features contribute more to early tokens, and late input features to late tokens?} Such temporal alignment would suggest that the model preserves sequential structure from input to output, which is particularly relevant for speech recognition where the temporal ordering of phonemes and words should be reflected in the generated transcription.

To enable comparison across samples with varying sequence lengths, we normalize both feature and token positions to percentages. For each modality $m \in \{A, V\}$, we group input features into $K$ temporal bins $\{\mathcal{F}_1^{(m)}, \ldots, \mathcal{F}_K^{(m)}\}$ and output tokens into $W$ temporal bins $\{\mathcal{T}_1, \ldots, \mathcal{T}_W\}$ (e.g., in Figure~\ref{fig:main_diagram} $K = W = 3$). We then compute, for each feature bin $k$, the proportion of its total contribution directed to each token bin $w$:
\begin{equation}
    H_{k,w}^{(m)} = \frac{\displaystyle\sum_{j \in \mathcal{F}_k^{(m)}} \sum_{t \in \mathcal{T}_w} |\phi_{j,t}|}{\displaystyle\sum_{w'=1}^{W} \sum_{j \in \mathcal{F}_k^{(m)}} \sum_{t \in \mathcal{T}_{w'}} |\phi_{j,t}|}.
    \label{eq:alignment_matrix}
\end{equation}
Each row of $\bm{H}^{(m)} \in \mathbb{R}^{K \times W}$ sums to one, representing how the contributions of each group are distributed across the generation. We analyze this at two levels of \textit{granularity}: a fine-grained heatmap visualization, and a grouped analysis that partitions features into early, middle, and late segments. Temporal alignment manifests as diagonal patterns, where early features contribute predominantly to early tokens and late features to late tokens, while uniform distributions indicate no temporal structure.

\begin{comment}
Finally, we investigate whether temporal correspondence exists between input feature positions and output token positions, i.e., whether early input features contribute more to early tokens, and late features to late tokens. Such alignment would suggest that the model preserves sequential structure from input to output.

We first visualize the full Shapley matrix $\bm{\Phi}^{(m)}$ for each modality $m \in \{A, V\}$ as a heatmap, where rows correspond to input features (ordered temporally) and columns correspond to generated tokens. This reveals fine-grained patterns of which input positions contribute to which output positions.

To quantify temporal alignment more directly, we partition features of each modality into $K$ temporal groups (e.g., early, middle, late) and compute the relative contribution of each group to each stage of token generation. For modality $m$, let $\mathcal{F}_k^{(m)}$ denote the $k$-th feature group and $\mathcal{T}_w$ the $w$-th token window. The contribution of feature group $k$ to token window $w$ is:
\begin{equation}
    A_{k,w}^{(m)} = \frac{\displaystyle\sum_{j \in \mathcal{F}_k^{(m)}} \sum_{t \in \mathcal{T}_w} |\phi_{j,t}|}{\displaystyle\sum_{k'=1}^{K} \sum_{j \in \mathcal{F}_{k'}^{(m)}} \sum_{t \in \mathcal{T}_w} |\phi_{j,t}|}.
    \label{eq:alignment_grouped}
\end{equation}
Strong diagonal patterns in $\bm{A}^{(m)}$ (e.g., early features contributing most to early tokens) indicate temporal alignment, while uniform distributions suggest no systematic temporal structure.
\end{comment}

\section{Experimental Setup}
\label{sec:experiment_details}

\begin{figure*}
     \centering
     \begin{subfigure}[b]{0.45\textwidth}
         \centering
         \includegraphics[width=\textwidth]{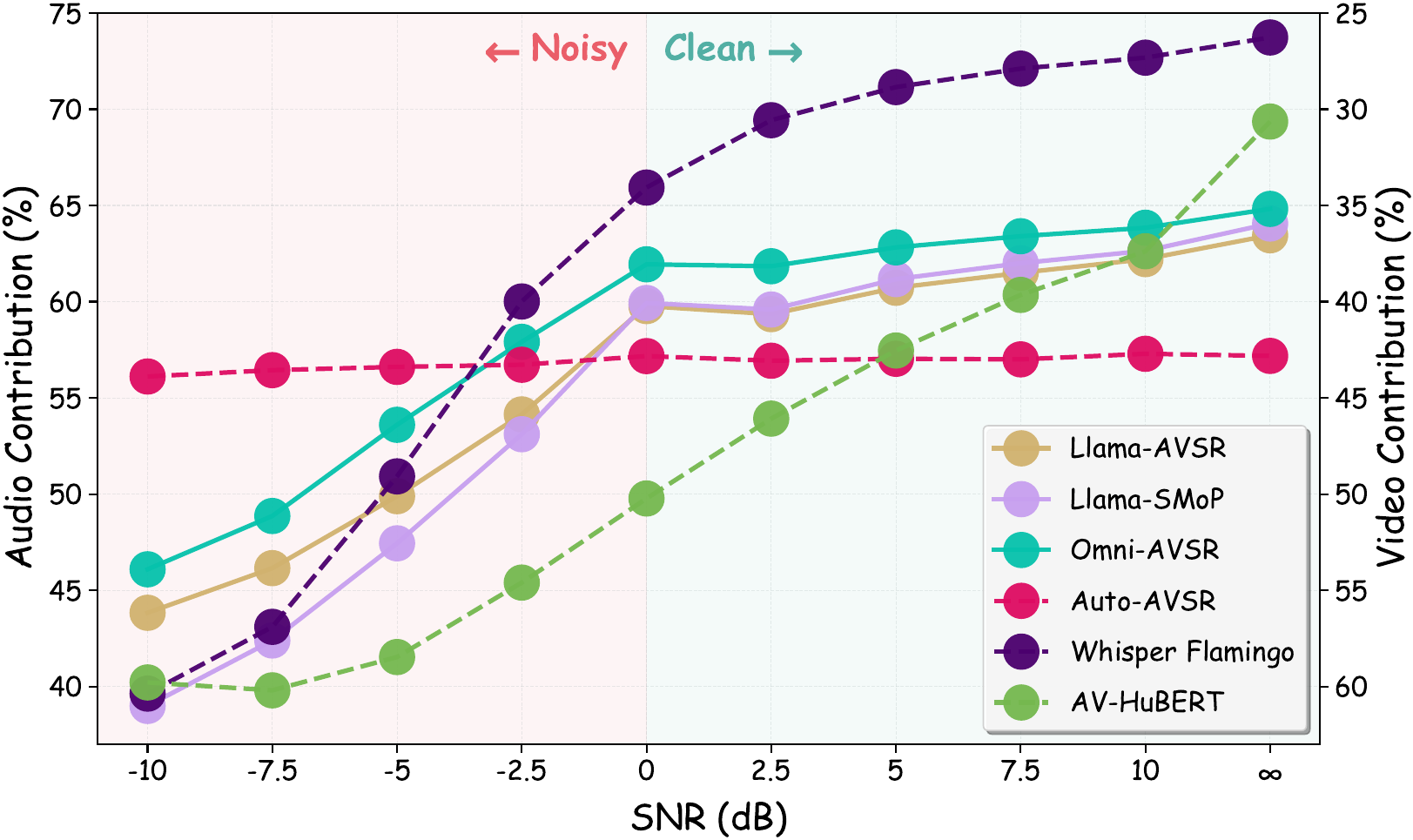}
     \end{subfigure}
     %\hfill
     \begin{subfigure}[b]{0.45\textwidth}
         \centering
         \includegraphics[width=\textwidth]{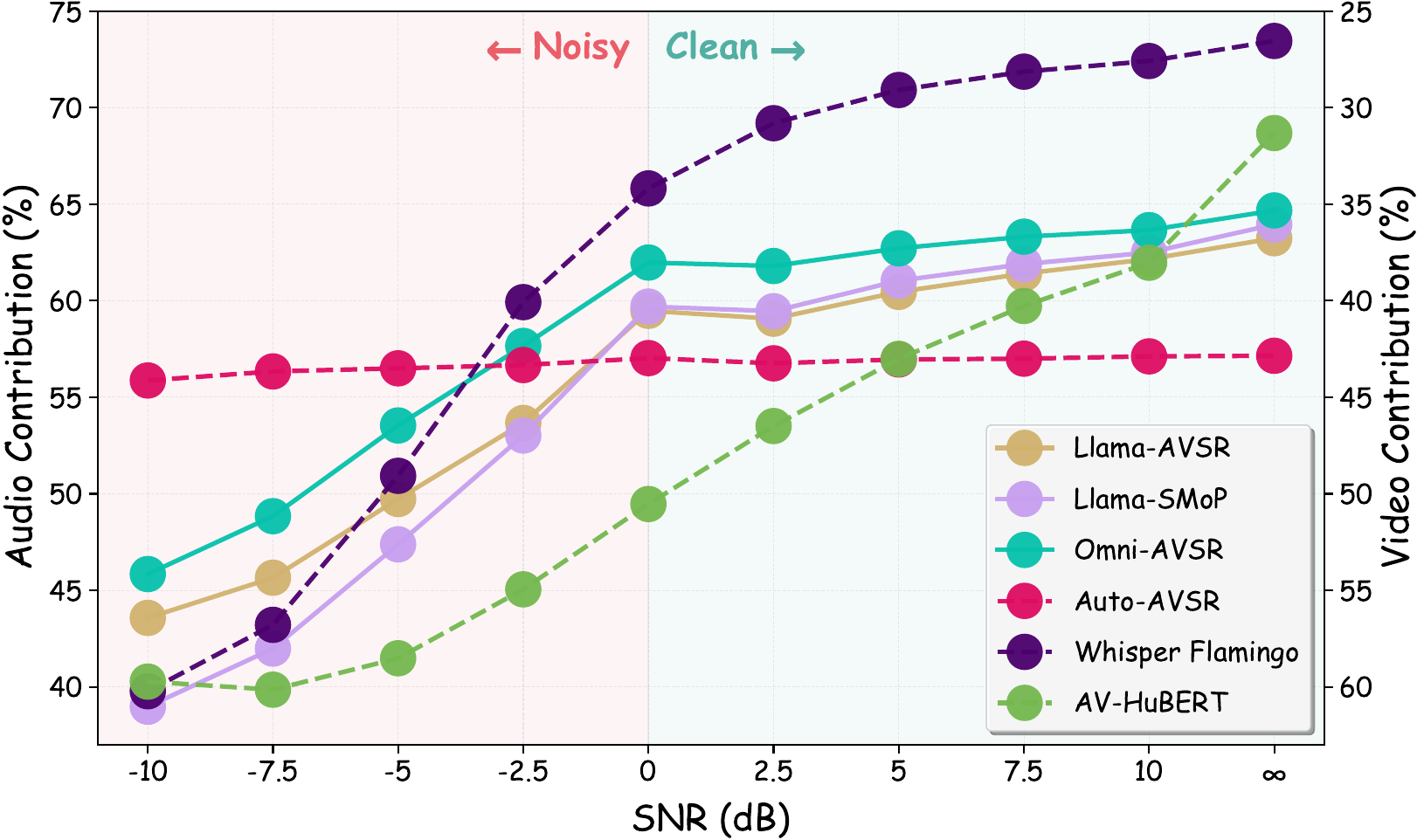}
     \end{subfigure}
    
    \caption{\textbf{(Left)}: Global audio/video contributions using \textbf{Permutation SHAP} for six AVSR models under varying acoustic conditions on the LRS3 dataset. \textbf{(Right)}: The same analysis using \textbf{Sampling SHAP}.}
    \label{fig:global_shap}
    \vspace{-0.5cm}
\end{figure*}
\subsection{Datasets} 
We conduct experiments on LRS2 \cite{son2017lip} and LRS3 \cite{afouras2018lrs3} datasets. LRS2 includes $225$ hours of footage from BBC programs. LRS3 contains $433$ hours of English video clips from TED talks.  

\subsection{Models} 
We analyze the modality contribution of six state-of-the-art AVSR models. We categorize them in two families: \textbf{(1)} \textbf{LLM}-based approaches, where the audio and visual inputs are processed by audio and visual pre-trained encoders, projected into the LLM space, and then fed to the LLM alongside the prompt tokens; \textbf{(2)} \textbf{cross-attention}-based, where the (attention-based) decoder processes the audio and visual embeddings via a cross-attention mechanism. We provide details about each model below. We point out that these models are used at inference time when we compute the Shapley values, and we follow the inference details described in each paper. %We will release the full code upon acceptance. 

\textbf{LLM-based Approaches}. \textbf{1)} \textit{Llama-AVSR}~\cite{llamaavsr} is the first multimodal LLM designed to perform the AVSR task. It generates text autoregressively conditioning on audio and video tokens obtained via modality-specific pre-trained encoders and linear projectors. \textbf{2)} \textit{Llama-SMoP}~\cite{cappellazzo2025scaling} extends Llama-AVSR by substituting the audio and video projectors with sparse mixture of experts modules to enhance audio-visual capabilities. We focus on the Joint-Experts, Joint-Router (JEJR) configuration which employs a shared router and shared experts to process both audio and video tokens. Owing to the fusion of information happening in this module, we expect Llama-SMoP to show a more balanced modality contribution with respect to Llama-AVSR. \textbf{3)} \textit{Omni-AVSR}~\cite{omniavsr} is a multimodal LLM that supports the tasks of ASR, VSR, and AVSR within a single model. In addition to this, Omni-AVSR is trained to support multi-granularities inference via matryoshka representation learning~\cite{cai2024matryoshka, kusupati2022matryoshka, cappellazzo2025mome, cappellazzo2025adaptive}. All three approaches use Whisper medium~\cite{radford2023robust} as audio encoder, AV-HuBERT Large~\cite{shi2022avhubert} as video encoder, and Llama 3.2-1B~\cite{dubey2024llama} as LLM backbone (we observed very similar trends when using Llama 3.2-3B). Additionally, these approaches downsample the audio and video features output of the encoders by a factor $4$x and $2$x, respectively, ensuring the number of audio and video features are the same (i.e., the frame rate is equal to $12.5$ Hz). 

\textbf{Cross-Attention-based Approaches}. \textbf{4)} \textit{AV-HuBERT}~\cite{shi2022avhubert} is a self-supervised framework for audio-visual speech representation learning. It learns joint audio-visual speech representations by predicting cluster assignments of masked audio-visual segments, using an offline clustering step followed by masked prediction training. For our experiments, we use the noise-augmented model pre-trained on LRS3 and VoxCeleb2~\cite{chung2018voxceleb2} and then fine-tuned on LRS3. \textbf{5)} \textit{Auto-AVSR}~\cite{autoavsr} processes the audio and visual streams through ResNet- and Conformer-based encoders and fused via MLP concatenation. The rest of the network consists of a projection layer and a Transformer decoder for joint CTC/attention training. For our experiments we use the model trained on $3448$ hours. \textbf{6)} \textit{Whisper-Flamingo}~\cite{rouditchenko2024whisper} extends the Whisper model~\cite{radford2023robust} by incorporating visual speech information through Flamingo-style~\cite{alayrac2022flamingo} gated cross-attention layers inserted into the Whisper decoder, enabling the model to attend to visual features extracted from lip movements. We use the model pre-trained on LRS3 and Voxceleb 2 with noise.

\subsection{Masking Features/Tokens}
Following~\cite{parcalabescu2023mm, parcalabescu2025decoders, morais2025investigating}, we mask excluded features/tokens by setting them to zero. While zero vectors may not perfectly correspond to marginalizing over absent features, our analysis focuses on relative modality attributions under consistent masking across all models and conditions, ensuring that observed trends reflect genuine differences in modality utilization rather than artifacts of the masking strategy. For LLM-based AVSR models, we mask both audio and video tokens after the projection layers (i.e., the tokens processed by the LLM). For cross-attention-based models, since AV-HuBERT and Auto-AVSR fuse the audio and visual tokens via an MLP layer and a transformer encoder, respectively, we mask the audio and video features before the fusion takes place. For Whisper-Flamingo, since no fusion happens due to the modality-specific cross-attention mechanism, we mask the cross-attended modality features before they are consumed by the decoder. Due to different encoder resolutions (Whisper: $50$ Hz audio vs. AV-HuBERT: $25$ fps video), we implement grouped audio masking to ensure fair SHAP attribution in Whisper-Flamingo. Each SHAP mask element controls two consecutive audio features ($0.04$s) but only one video feature ($0.04$s), equalizing the temporal coverage removed by masking. This prevents systematic bias toward the higher-resolution audio modality. %and ensures attribution percentages reflect true modality importance. 

\section{Results}

\subsection{Global Modality Contributions Across SNR Levels}
We analyze how audio and visual contributions vary with acoustic conditions using \texttt{Global} \texttt{SHAP}. Figure~\ref{fig:global_shap} presents the audio/visual contribution for all six AVSR models across SNR levels ranging from $-10$~dB (severely degraded audio) to clean speech ($\infty$) on the LRS3 dataset. Following the types of noise reported in the corresponding papers, we use babble noise sampled from the NOISEX dataset~\cite{varga1993assessment} for the LLM-based methods and Auto-AVSR, whereas for Whisper-Flamingo and AV-HuBERT the babble noise is sampled from the MUSAN dataset~\cite{snyder2015musan}. %We evaluate both Permutation SHAP and Sampling SHAP to assess the robustness of our approximations.

\textbf{Consistent Approximations}. Both Permutation SHAP (left) and Sampling SHAP (right) yield similar results across all models and SNR conditions, confirming that our approximations with $M=2000$ coalitions provide stable and reliable modality attributions (reducing $M$ yielded higher variance, particularly at low SNR). Consequently, all subsequent results are presented using Permutation SHAP, which is computationally faster.

\textbf{Dynamic Modality Adaptation}. The results in Figure~\ref{fig:global_shap} reveal that most AVSR models dynamically adjust their reliance on audio and visual modalities based on acoustic conditions. At low SNR ($-10$~dB), where audio is severely corrupted by noise, models shift toward greater visual reliance, with audio contributions dropping to $39$-$46$\% for most models. As SNR increases and audio quality improves, models progressively increase their audio dependence, reaching $63$-$73$\% in clean conditions. This adaptive behavior demonstrates that these models have learned to leverage the complementary nature of audio and visual information. However, we note that \textit{all models maintain substantial audio reliance even under severe noise conditions}, where one might expect near-complete dependence on the visual modality. We conjecture that this is attributable to the attention mechanism of the decoder, which continues to attend heavily to audio features regardless of their quality. This suggests that task-specific adaptation strategies may be needed to encourage more appropriate modality weighting under adverse acoustic conditions.

\begin{table}[t]
\renewcommand{\tabcolsep}{1.2mm}
\centering
    \caption{Global \textbf{audio} contributions for four AVSR models under varying acoustic conditions on the LRS2 dataset.}
%\resizebox{0.6\linewidth}{!}{
\begin{tabular}{lcccccc}

\toprule
\multirow{2}{*}{\textbf{Method}} & \multicolumn{6}{c}{\cellcolor{teagreen}\textbf{SNR (dB)}}\\
      \cmidrule(rl){2-7} & $\infty$ & 10 & 5 & 0 & -5 & -10\\

\midrule
Llama-AVSR~\cite{llamaavsr} &65.4 &64.5 &63.3 &62.2 &52.7 & 46.0 \\
Omni-AVSR~\cite{omniavsr} &59.9 &59.9 &58.2 &57.3 &48.8 &42.7 \\
Auto-AVSR~\cite{autoavsr} &56.5 &56.4 &56.4 &56.3 &55.9 &55.3 \\
Whisper-Flamingo~\cite{rouditchenko2024whisper} &94.7 &94.5 &94.2 &93.1 &90.2 &88.3 \\
\bottomrule
 \end{tabular}%}
\label{tab:lrs2}
\vspace{-0.5cm}
\end{table}

\textbf{Model-Specific Patterns}. We observe substantial variation across architectures. Whisper-Flamingo and AV-HuBERT exhibit the widest adaptation ranges, both shifting by approximately $30$-$34$ percentage points between $-10$~dB and clean speech. However, the shape of their adaptation differs markedly: Whisper-Flamingo transitions rapidly in the negative SNR range (from $40$\% at $-10$~dB to $66$\% at $0$ dB) before plateauing in clean conditions, whereas AV-HuBERT shifts more gradually across the entire SNR range. The LLM-based models show moderate adaptation of $19$-$25$ points, with similar overall trajectories. Among them, Llama-SMoP exhibits the greatest visual reliance under extreme noise ($39$\% audio at $-10$~dB), likely due to its mixture-of-experts routing, while Omni-AVSR maintains comparatively higher audio reliance across all conditions, consistent with its multi-task training on ASR, VSR, and AVSR which reinforces audio bias. Strikingly, Auto-AVSR exhibits almost no adaptation, maintaining nearly constant audio contribution (around $57$\%) across all SNR levels. This is consistent with its MLP-based fusion, which applies fixed learned weights regardless of input quality, unlike attention-based mechanisms (i.e., AV-HuBERT) that can dynamically modulate modality reliance.

\textbf{Additional Results on LRS2}. Table~\ref{tab:lrs2} reports global audio SHAP contributions on LRS2 (for models with available LRS2 checkpoints). The results are consistent with LRS3: Llama-AVSR and Omni-AVSR show adaptive behavior, with audio contributions decreasing from $65.4\%$ and $59.9\%$ in clean conditions to $46.0$\% and $42.7$\% at $-10$~dB, respectively. Auto-AVSR again maintains a flat profile at approximately $56$\%. Whisper-Flamingo exhibits a stronger audio bias on LRS2 ($88$-$95$\%) than LRS3, likely because its frozen AV-HuBERT video encoder was fine-tuned on LRS3 and not LRS2, causing domain mismatch amplified by the smaller LRS2 training set. Overall, these consistent patterns suggest that modality behaviors are intrinsic to each model's architecture rather than dataset-specific. All subsequent results are reported on LRS3.

%Table~\ref{tab:lrs2} reports the Global SHAP audio contributions on the LRS2 dataset (we consider models which provide checkpoints trained on LRS2). The results are consistent with our findings on LRS3: Llama-AVSR and Omni-AVSR show adaptive modality behavior, with audio contributions decreasing from $65.4$\% and $59.9$\% in clean conditions to $46.0$\% and $42.7$\% at $-10$~dB, respectively. Auto-AVSR again exhibits a flat profile, maintaining approximately $56$\% audio contribution regardless of acoustic conditions. Whisper-Flamingo exhibits strong audio reliance on both datasets due to its Whisper backbone, but this bias is even more pronounced on LRS2 ($88$-$95$\%) compared to LRS3. We surmise this is because Whisper-Flamingo keeps the video encoder AV-HuBERT frozen, which is fine-tuned on LRS3 and not on LRS2, thus causing domain mismatch issues. Combined with the smaller size of the LRS2 training set, this further amplifies the model's dependence on audio. Overall, these consistent patterns across datasets suggest that the observed modality behaviors are intrinsic to each model's architecture rather than dataset-specific artifacts. In the next sections we will report the results obtained on LRS3.

\begin{tcolorbox}[enhanced, drop fuzzy shadow, colback=darkmaroon!15, colframe=darkmaroon, title=Finding 1: Adaptive Modality Shift with Persistent Audio Bias, fonttitle=\bfseries] AVSR models tend to shift toward visual reliance as noise increases, yet maintain high audio contributions even under severe degradation. \end{tcolorbox}

\begin{figure}[t]
    \centering
    \includegraphics[width=\linewidth]{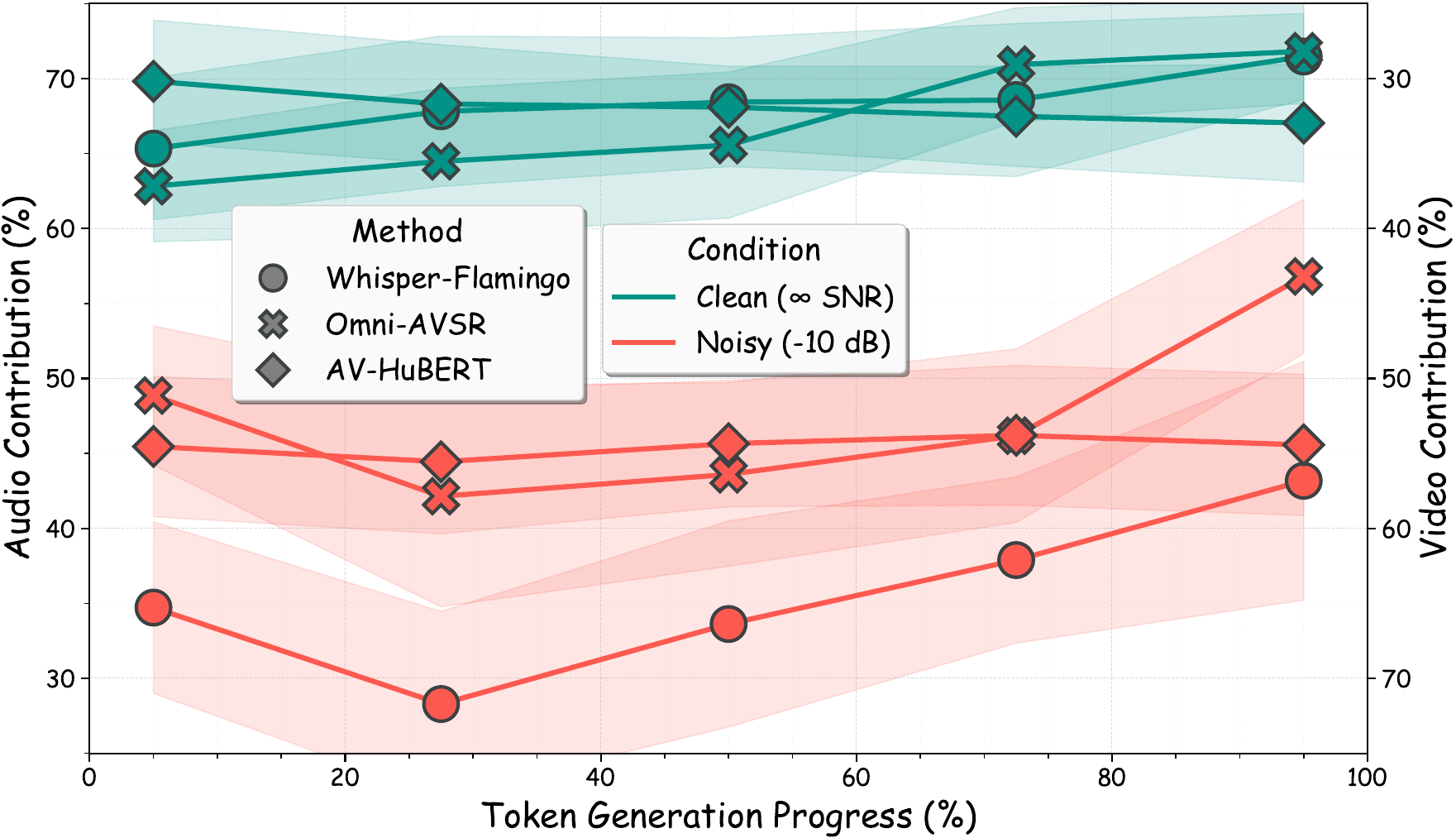}
    \caption{\texttt{Generative} \texttt{SHAP} analysis showing modality contributions as a function of token generation progress (\%). \hlclean{Clean} and \hlnoisy{noisy} ($-10$~dB) conditions are compared.} %Shaded regions indicate standard deviation across the 20 longest utterances.}
    \label{fig:generative_shap}
    \vspace{-0.7cm}
\end{figure}

\subsection{Modality Contributions During Generation}
We estudy how modality contributions evolve throughout the token generation process using \texttt{Generative} \texttt{SHAP}. Figure~\ref{fig:generative_shap} presents the audio/video contributions across generation progress for Whisper-Flamingo, Omni-AVSR (the remaining LLM-based models showed similar trends), and AV-HuBERT under clean and noisy conditions. To better capture temporal dynamics, we focus on the $20$ longest utterances (around $6$ seconds each), 
which ensures sufficient token length for stable windowed estimates ($W = 5$) and more fine-grained contribution trajectories.

\textbf{Whisper-Flamingo and Omni-AVSR: Increasing Audio Reliance.} In clean conditions, both models show notable increases in audio contribution as generation progresses, from $65$\% to $71$\% for Whisper-Flamingo and $63$\% to $72$\% for Omni-AVSR. Under noisy conditions, a distinctive U-shaped pattern emerges: both models begin with moderate audio reliance, dip to their lowest around mid-generation, then increase substantially toward the end. This suggests that models initially depend on visual information when facing degraded audio, but progressively recover audio utilization as accumulated linguistic context aids interpretation of the noisy signal.

\textbf{AV-HuBERT: Stable Modality Balance.} AV-HuBERT exhibits markedly different behavior. In both clean and noisy conditions, audio contribution remains remarkably stable, with less than $3$ points variation throughout generation (the audio contribution slightly decreases). This consistency likely reflects its self-supervised pre-training on masked audio-visual prediction, which encourages balanced multimodal integration.

\begin{tcolorbox}[enhanced, drop fuzzy shadow, colback=darkmaroon!15, colframe=darkmaroon, title=Finding 2: Dynamic Modality Shift During Generation, fonttitle=\bfseries] Whisper-Flamingo and Omni-AVSR progressively increase audio reliance during generation, while AV-HuBERT maintains stable modality balance throughout. \end{tcolorbox}

\begin{figure}
     \centering
     \begin{subfigure}[b]{0.23\textwidth}
         \centering
         \includegraphics[width=\textwidth]{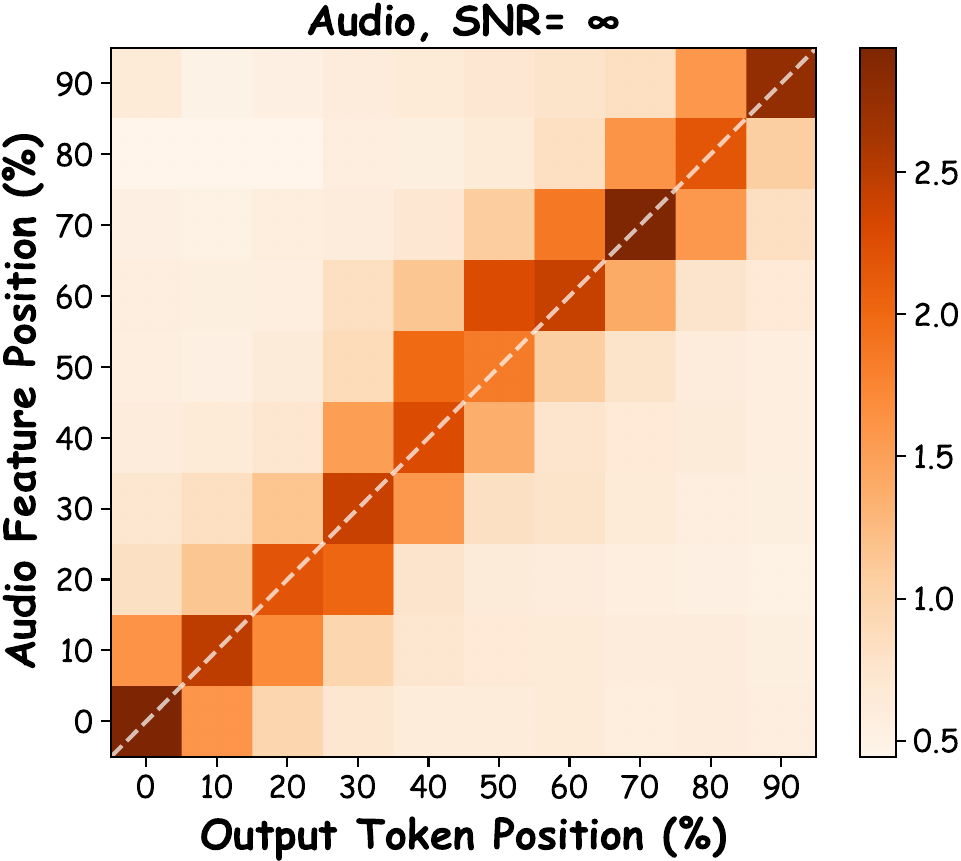}
     \end{subfigure}
     \begin{subfigure}[b]{0.23\textwidth}
         \centering
         \includegraphics[width=\textwidth]{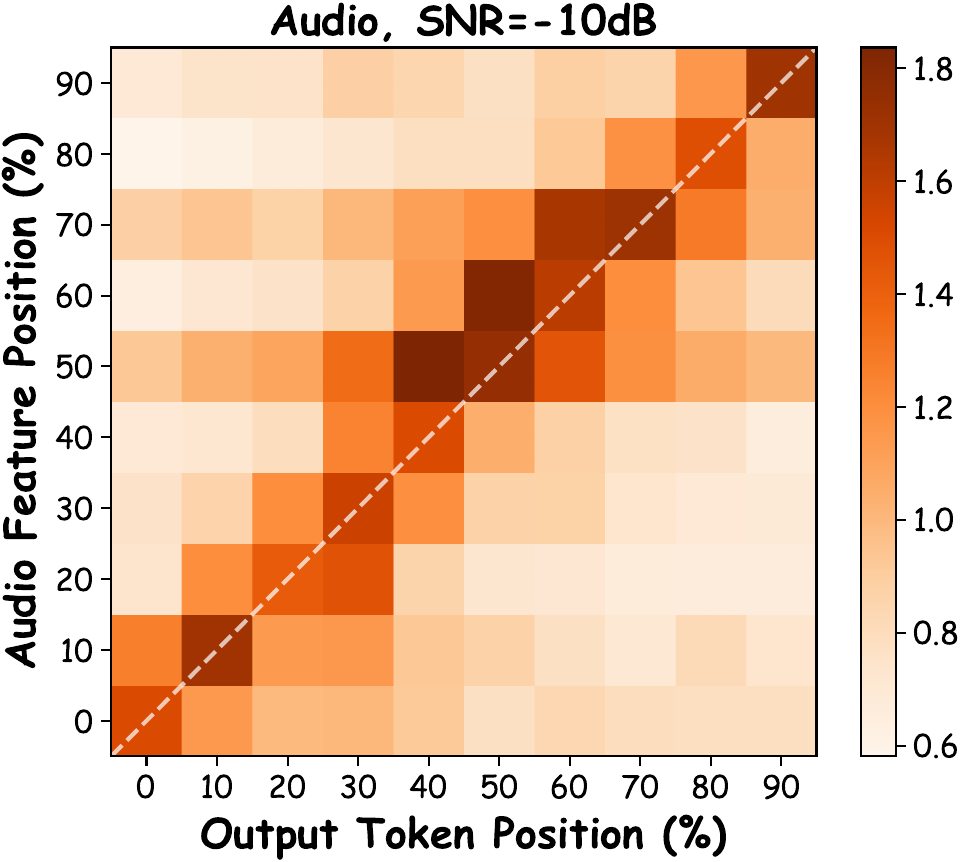}
     \end{subfigure}

     \begin{subfigure}[b]{0.23\textwidth}
         \centering
         \includegraphics[width=\textwidth]{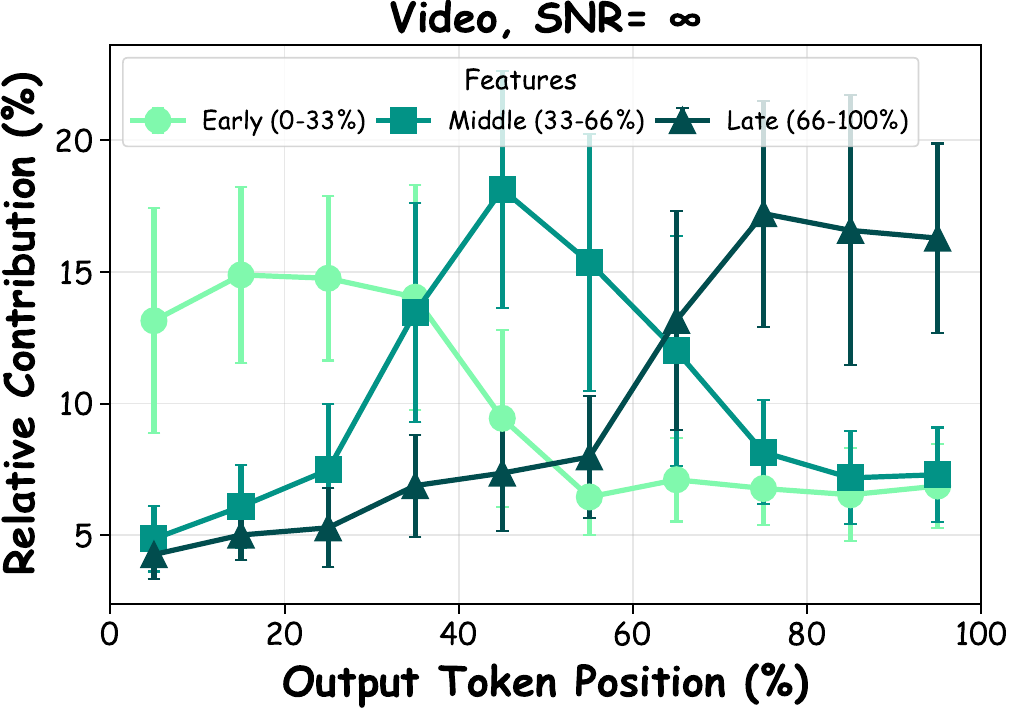}
     \end{subfigure}
     \begin{subfigure}[b]{0.23\textwidth}
         \centering
         \includegraphics[width=\textwidth]{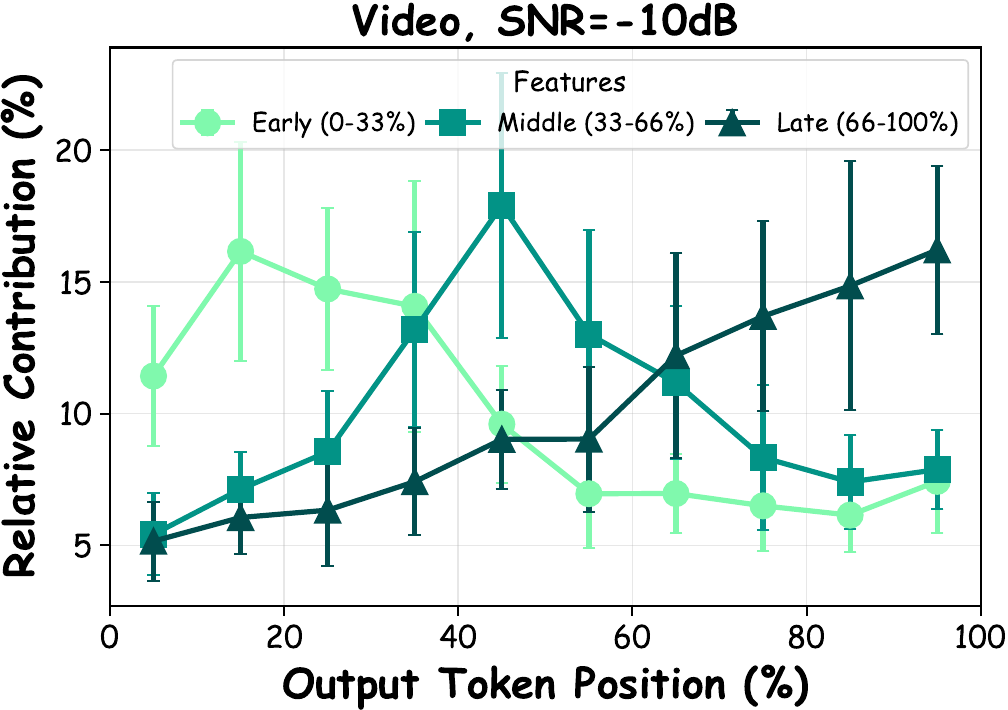}
     \end{subfigure}
    
    \caption{\texttt{Temporal} \texttt{Alignment} \texttt{SHAP} for AV-HuBERT. \textbf{Top}: audio feature heatmaps under clean \textbf{(left)} and noisy \textbf{(right)} conditions. \textbf{Bottom}: grouped video feature analysis (early/middle/late) under clean \textbf{(left)} and noisy \textbf{(right)} conditions.}
    \label{fig:heatmaps}
    \vspace{-0.6cm}
\end{figure}

\subsection{Input-Output Temporal Alignment Analysis}

We study whether AVSR models preserve temporal structure by examining the correspondence between input feature positions and output token positions using our proposed \texttt{Temporal} \texttt{Alignment} \texttt{SHAP} analysis. We observed consistent patterns across all evaluated models; here we present AV-HuBERT as a representative example, analyzing both clean and noisy conditions using fine-grained heatmaps ($K = W = 10$) and grouped feature analysis (early/middle/late, $K = 3$, $W = 10$).

\textbf{Strong Temporal Alignment.} The heatmaps in the upper part of Figure~\ref{fig:heatmaps} show audio feature contributions under clean (diagonal alignment score = $2.90$) and noisy conditions (diagonal alignment score = $1.70$), revealing clear \textit{diagonal} patterns: early features contribute predominantly to early tokens, middle features to middle tokens, and late features to late tokens. The grouped analysis in the lower part of Figure~\ref{fig:heatmaps}, shown for video features, confirms this: early, middle, and late feature groups show well-separated trajectories that peak at corresponding generation stages. Similar patterns are observed for video heatmaps and audio grouped plots, omitted for space. Crucially, both modalities maintain this temporal structure, indicating that audio and visual streams independently preserve sequential correspondence with the output rather than one modality dominating the alignment. Remarkably, this holds even under severe noise ($-10$~dB): although the diagonal pattern broadens and grouped trajectories become more overlapping, neither modality's alignment collapses when input quality degrades.

\begin{tcolorbox}[enhanced, drop fuzzy shadow, colback=darkmaroon!15, colframe=darkmaroon, title=Finding 3: Robust Temporal Alignment, fonttitle=\bfseries] Both audio and visual modalities independently maintain temporal correspondence between input features and output tokens, even under severe acoustic noise. \end{tcolorbox}

\begin{figure}
     \centering
     \begin{subfigure}[b]{0.45\textwidth}
         \centering
         \includegraphics[width=\textwidth]{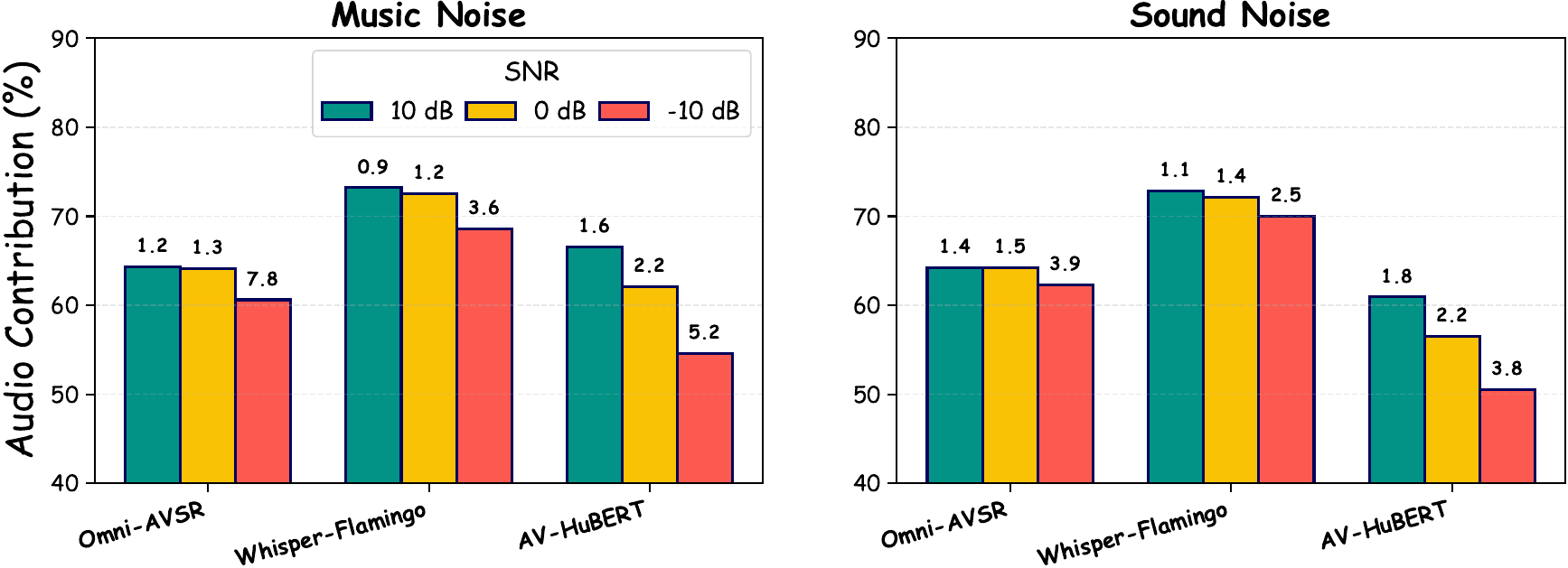}
     \end{subfigure}
     \begin{subfigure}[b]{0.45\textwidth}
         \centering
         \includegraphics[width=\textwidth]{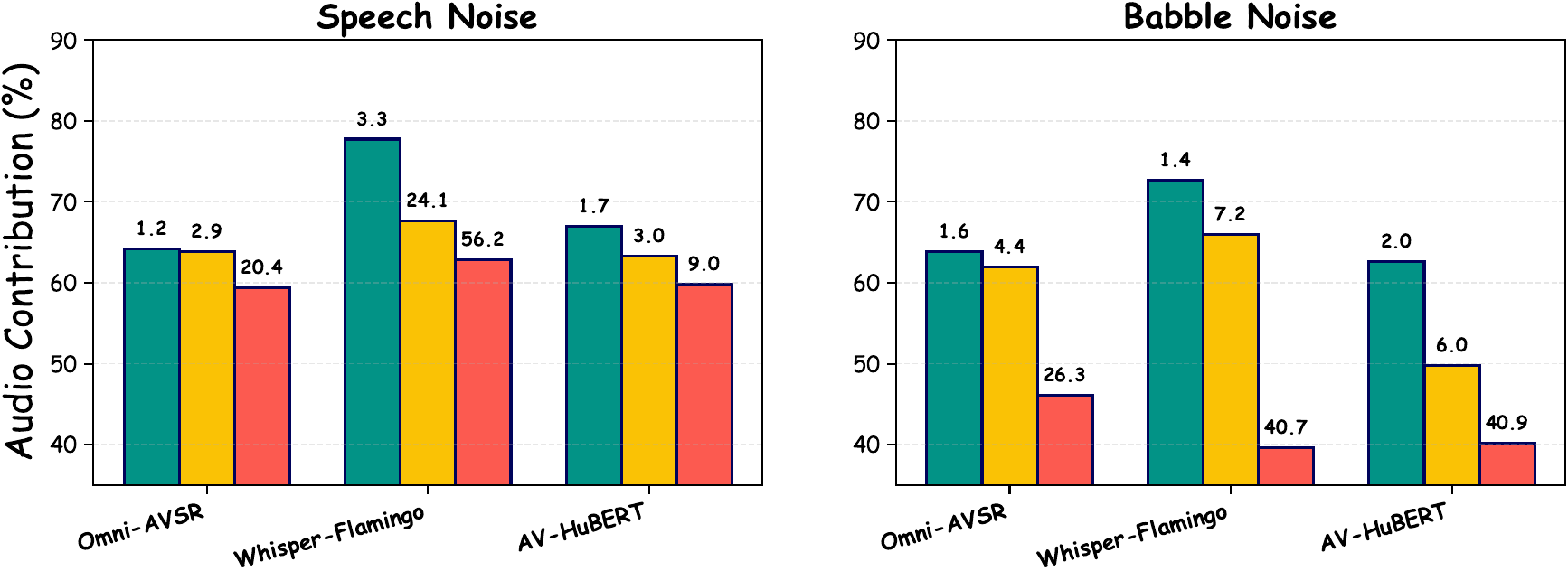}
     \end{subfigure}
    \caption{Global SHAP audio contribution across acoustic noise types. Numbers above bars indicate the achieved WER (\%).}
    \label{fig:different_noise}
    \vspace{-0.6cm}
\end{figure}

\begin{figure*}
     \centering
     \begin{subfigure}[b]{0.16\textwidth}
         \centering
         \includegraphics[width=\textwidth]{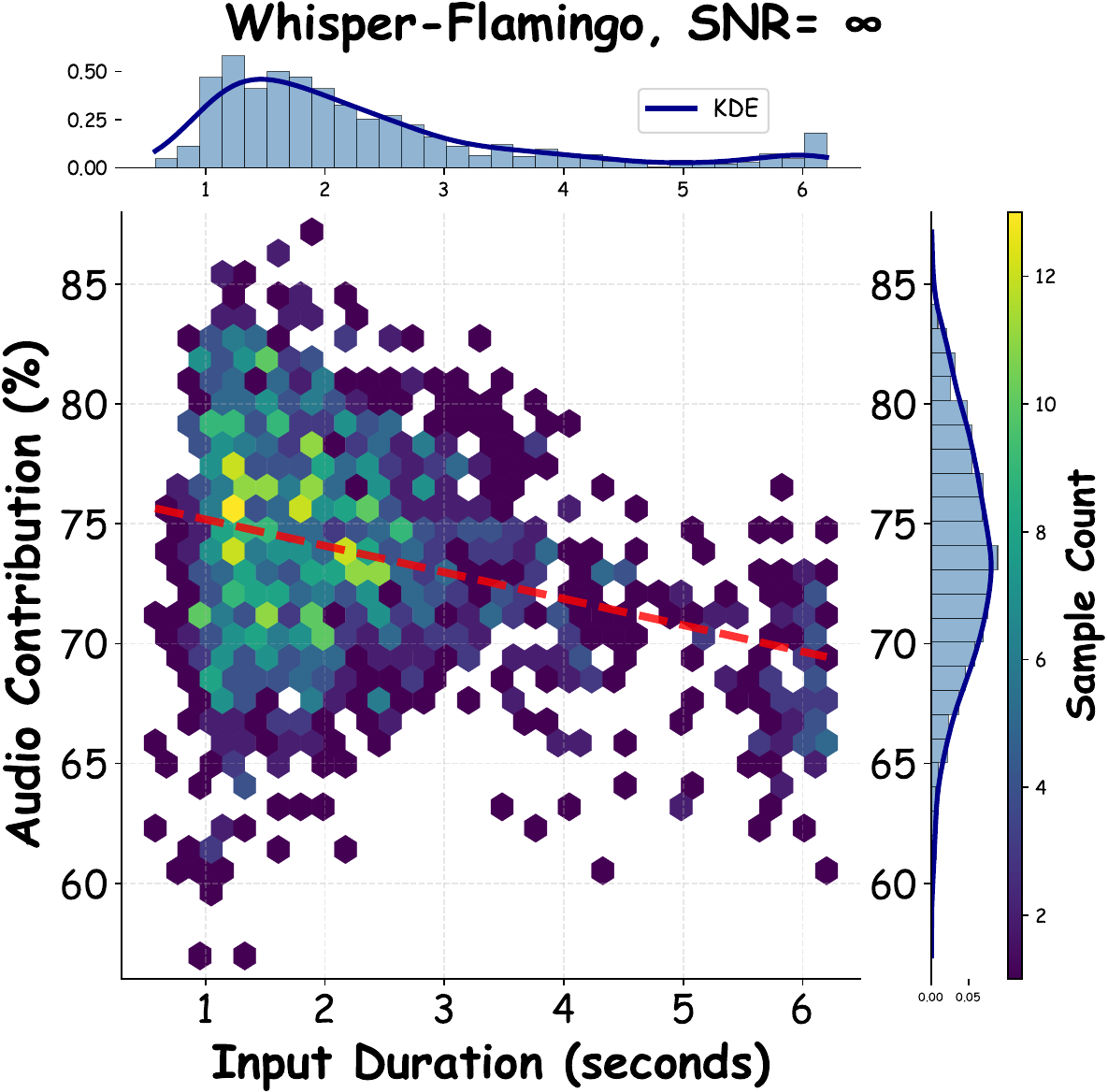}
     \end{subfigure}
     \begin{subfigure}[b]{0.16\textwidth}
         \centering
         \includegraphics[width=\textwidth]{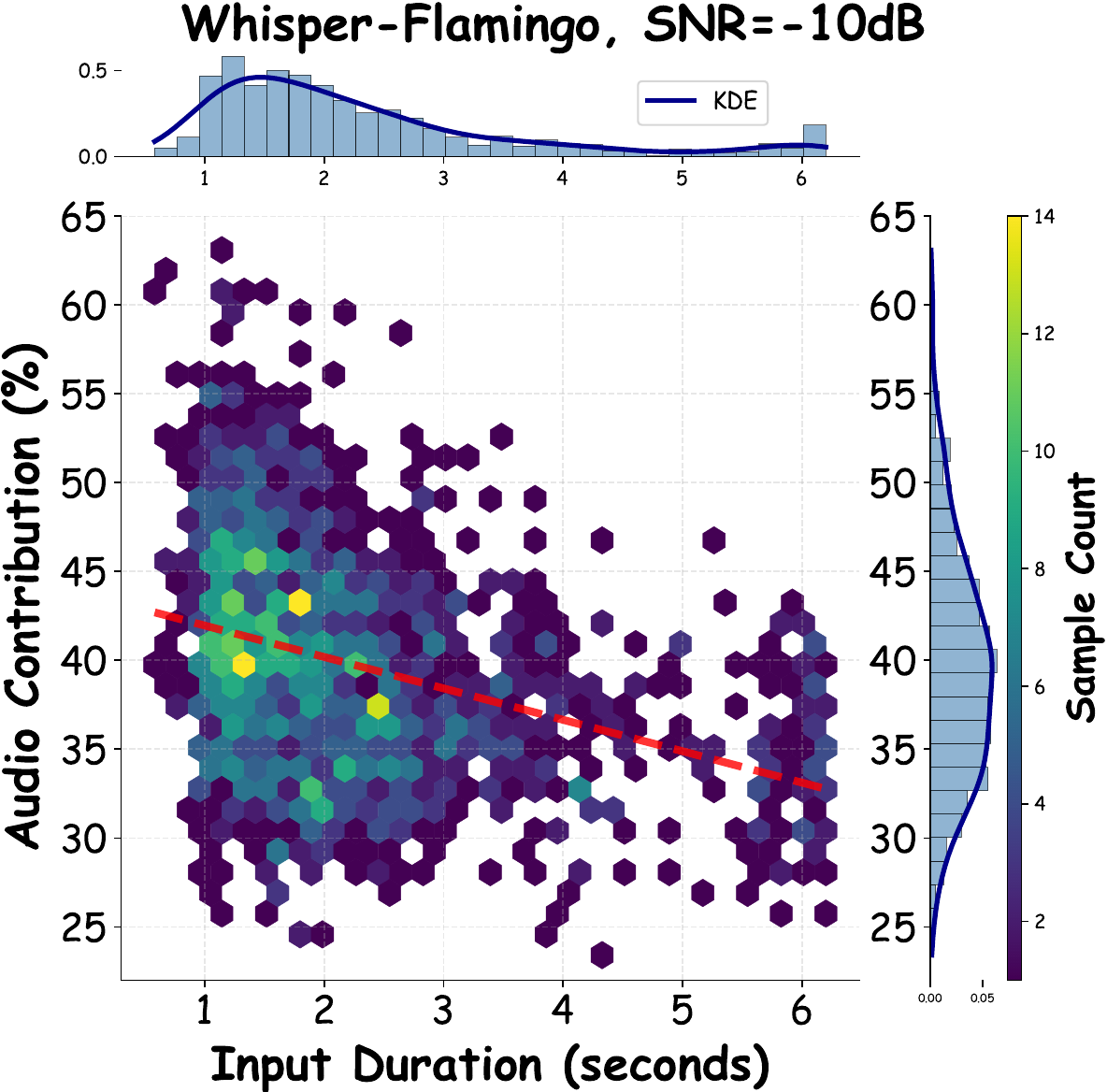}
     \end{subfigure}
     \begin{subfigure}[b]{0.16\textwidth}
         \centering
         \includegraphics[width=\textwidth]{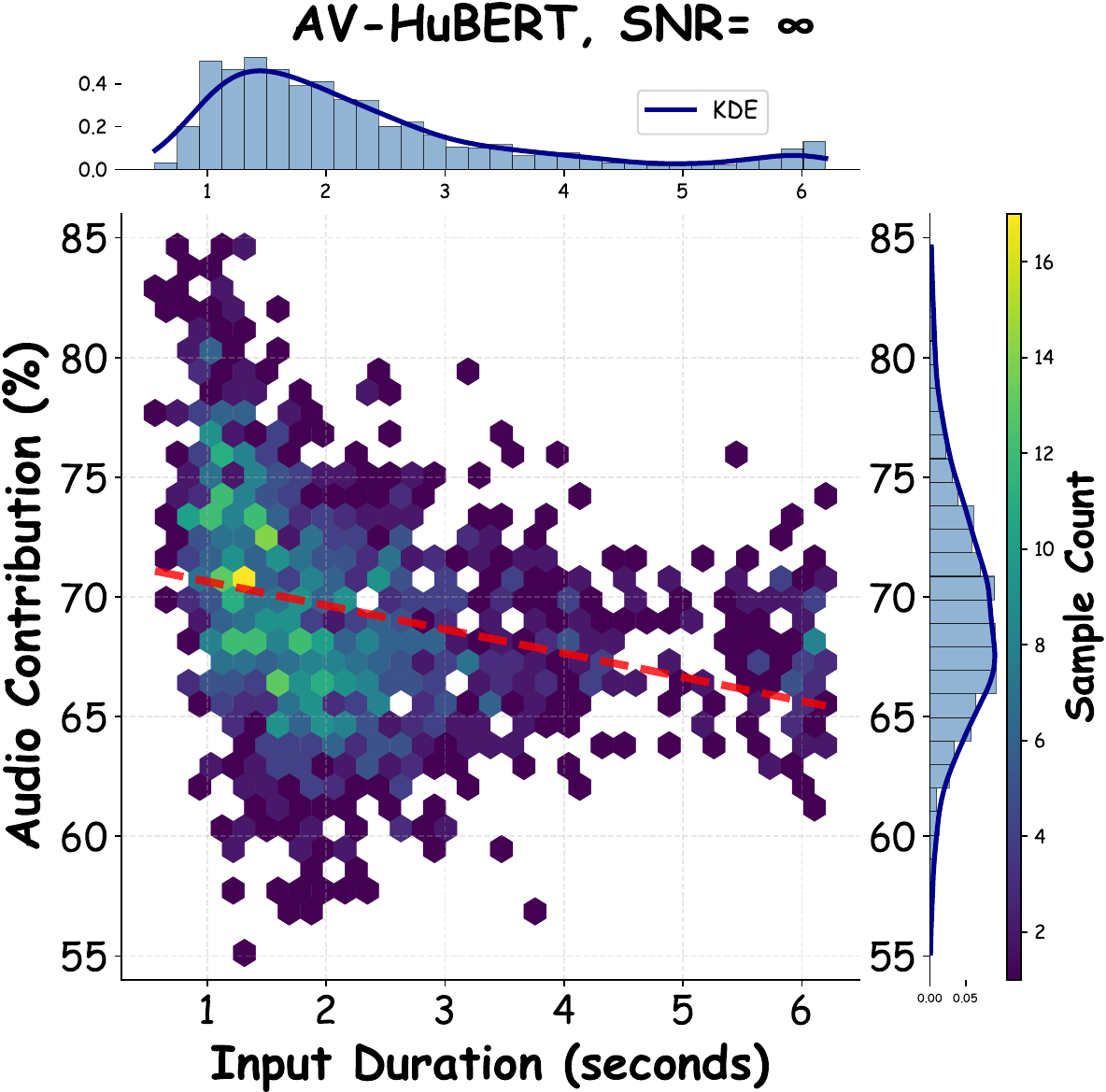}
     \end{subfigure}
     \begin{subfigure}[b]{0.16\textwidth}
         \centering
         \includegraphics[width=\textwidth]{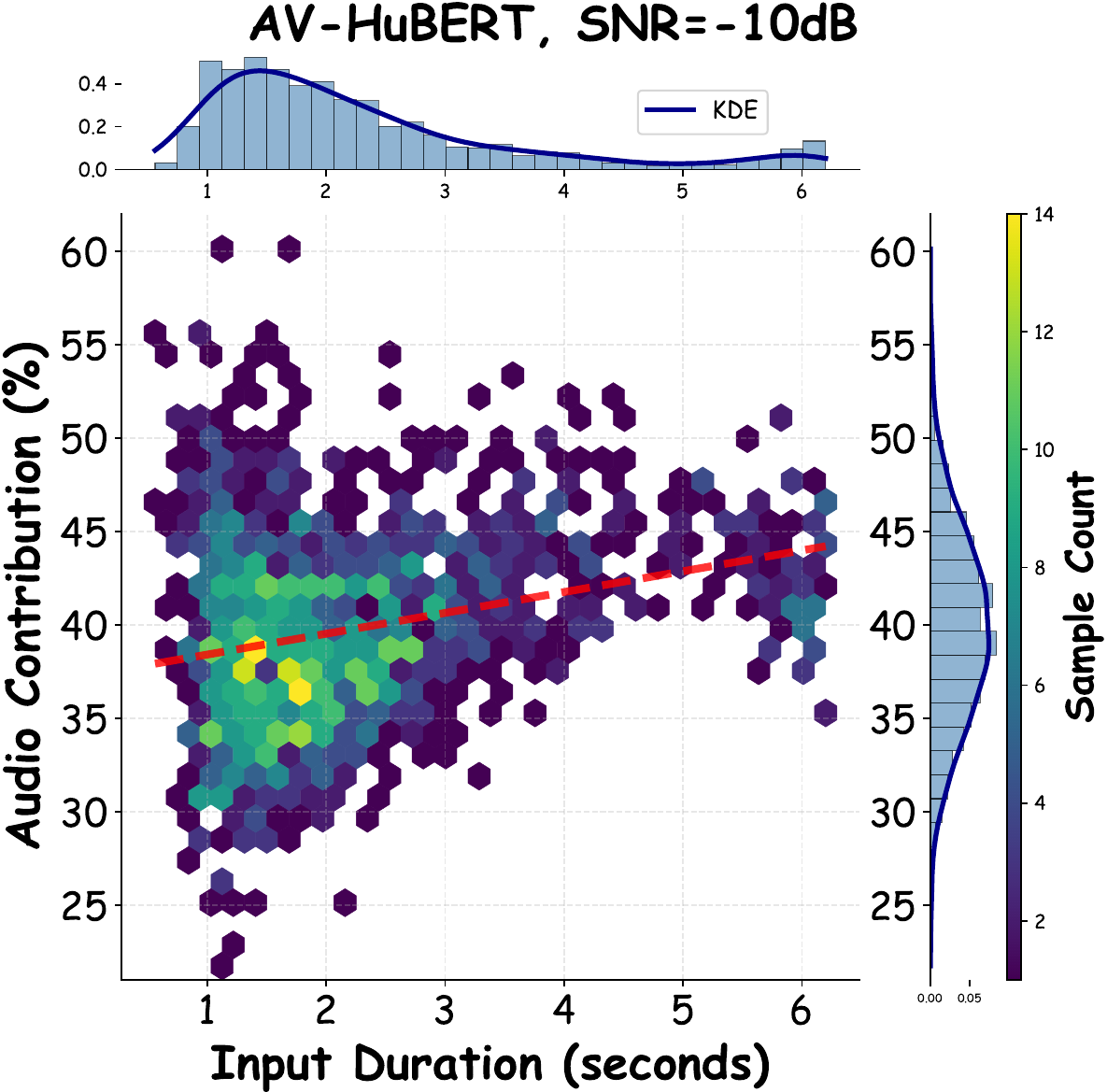}
     \end{subfigure}
     \begin{subfigure}[b]{0.16\textwidth}
         \centering
         \includegraphics[width=\textwidth]{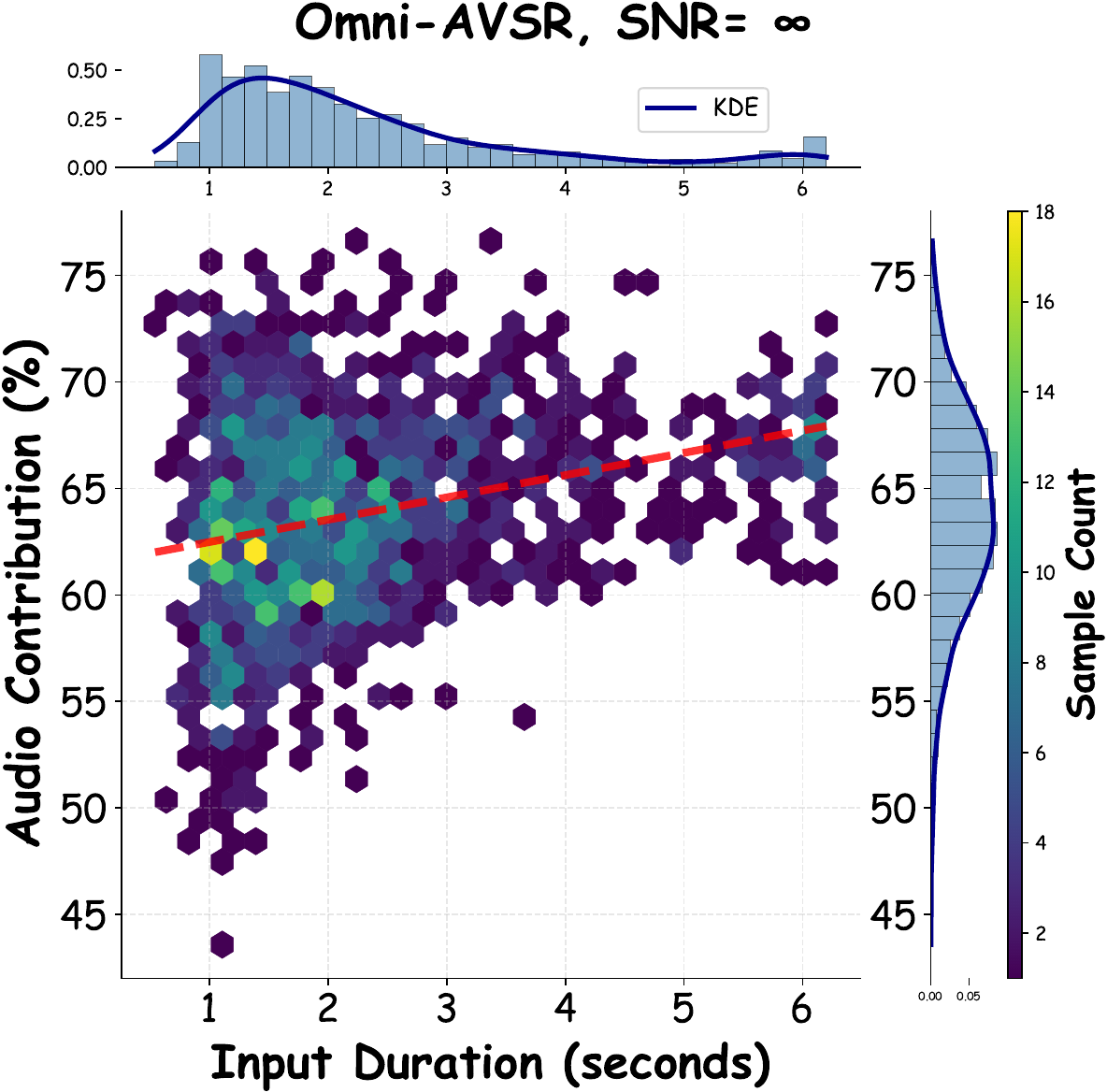}
     \end{subfigure}
     \begin{subfigure}[b]{0.16\textwidth}
         \centering
         \includegraphics[width=\textwidth]{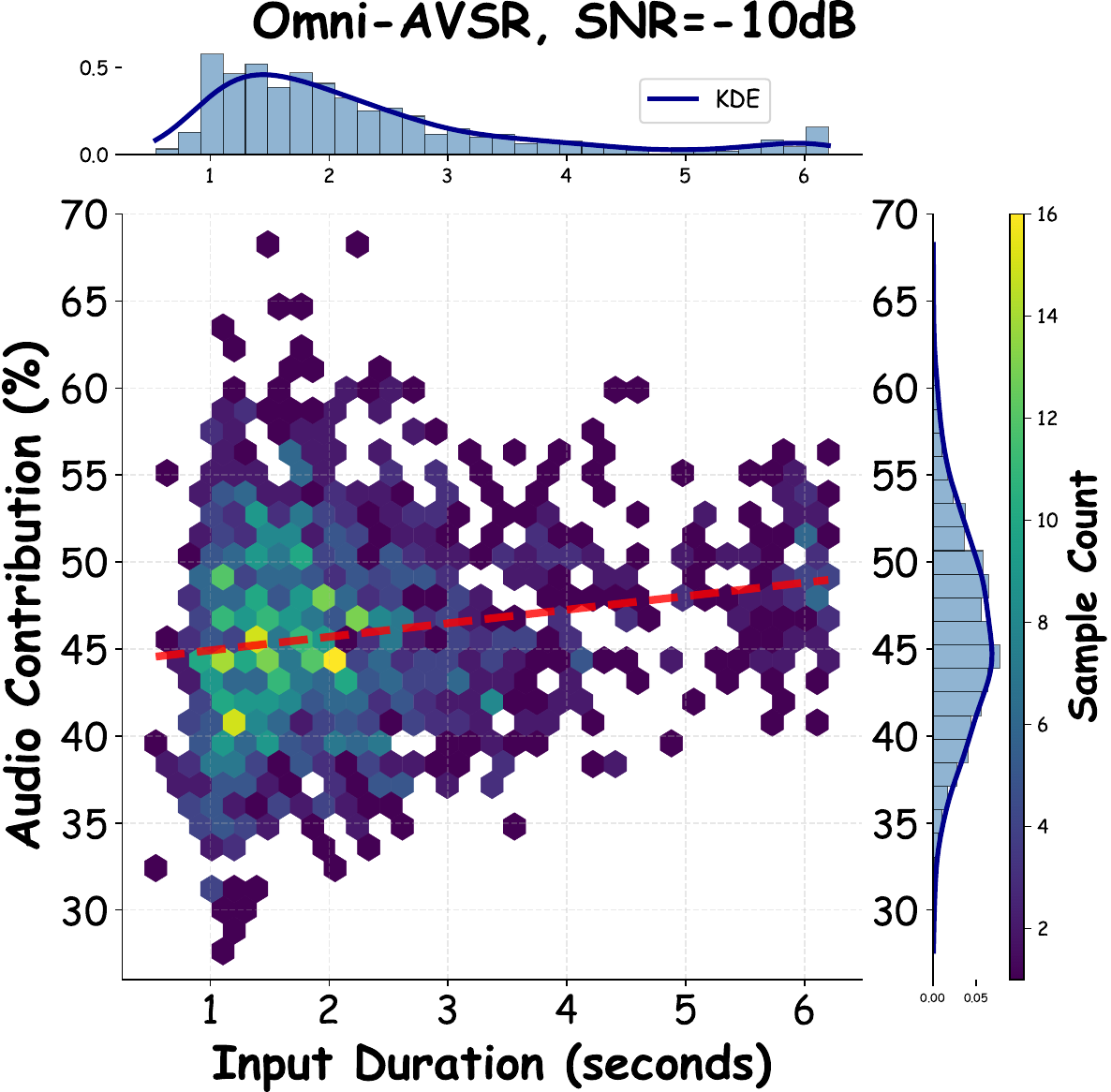}
     \end{subfigure}
    
    \caption{Global audio SHAP contributions as a function of input duration (seconds) for Omni-AVSR, Whisper-Flamingo, and AV-HuBERT in clean and noisy conditions. Marginal distributions are shown on top and right axes.}
    \label{fig:duration}
    \vspace{-0.3cm}
\end{figure*}

\subsection{Impact of Noise Type on Modality Contributions}
We examine how different noise types affect modality contributions. Figure~\ref{fig:different_noise} presents global audio contributions for babble noise and three noise categories from the MUSAN dataset~\cite{snyder2015musan}: \textit{music}, \textit{environmental sound}, and \textit{speech}, at various SNR levels. WER values ($\%$) are annotated above each bar.

\textbf{Moderate Shift Toward Visual Modality.} Compared to babble noise, these noise types induce a smaller shift toward visual reliance. This is also reflected in the relatively lower WER values: models achieve WERs in the range $2.6$-$7.8$ for music and sound noise compared to $26.3$-$40.9$ for babble noise in the $-10$~dB setting. Notably, Omni-AVSR was trained exclusively on babble noise and has never encountered MUSAN noise types, yet still demonstrates reasonable robustness. Consistent with our earlier findings, AV-HuBERT exhibits the most elastic modality adaptation, showing the largest shift from audio toward visual reliance as SNR decreases across all noise types. 

\begin{tcolorbox}[enhanced, drop fuzzy shadow, colback=darkmaroon!15, colframe=darkmaroon, title=Finding 4: Noise-Dependent Modality Adaptation, fonttitle=\bfseries] The degree of visual shift depends on noise type and severity: more challenging acoustic conditions induce greater reliance on visual information. \end{tcolorbox}

\subsection{Effect of Input Duration on Modality Contributions} 
We analyze how utterance duration affects modality contributions by examining global audio SHAP values across inputs of varying lengths (we take into account all $1321$ utterances from the LRS3 test set). Figure~\ref{fig:duration} presents hexbin density plots showing the relationship between input duration and audio contributions, with marginal distributions and trend lines (video contributions follow directly as V-SHAP $= 1 -$ A-SHAP).

\textbf{Model-Dependent Duration Effects.} The relationship between utterance duration and modality balance varies across architectures. Whisper-Flamingo shows decreasing audio contribution with duration in both clean ($75$\%$\,\to\,$$69$\%) and noisy ($42$\%$\,\to\,$$33$\%) conditions, with the effect amplified under noise. We attribute this to its gated cross-attention mechanism: as utterances grow longer, the decoder accesses richer visual context through cross-attention, and under noise, cumulative audio degradation further tilts the balance toward vision. AV-HuBERT exhibits a similar negative trend in clean conditions but, notably, reverses under noise, where longer utterances slightly favor audio. We hypothesize that AV-HuBERT's self-supervised pre-training on masked audio-visual prediction enables it to aggregate residual spectral cues from longer degraded audio sequences, while visual information offers diminishing returns due to the inherent redundancy of lip movements over time. Omni-AVSR shows a mild positive trend in both conditions: as an LLM-based model where all tokens attend to each other via self-attention, longer sequences allow audio tokens to mutually reinforce, regardless of acoustic conditions.

\begin{tcolorbox}[enhanced, drop fuzzy shadow, colback=darkmaroon!15, colframe=darkmaroon, title=Finding 5: Model-Dependent Duration Effects, fonttitle=\bfseries]
The relationship between utterance duration and modality balance is architecture-dependent, with no universal trend across models or conditions.
\end{tcolorbox}

\subsection{Does Recognition Difficulty Affect Modality Balance?}

\begin{figure}
     \centering
     \begin{subfigure}[b]{0.35\textwidth}
         \centering
         \includegraphics[width=\textwidth]{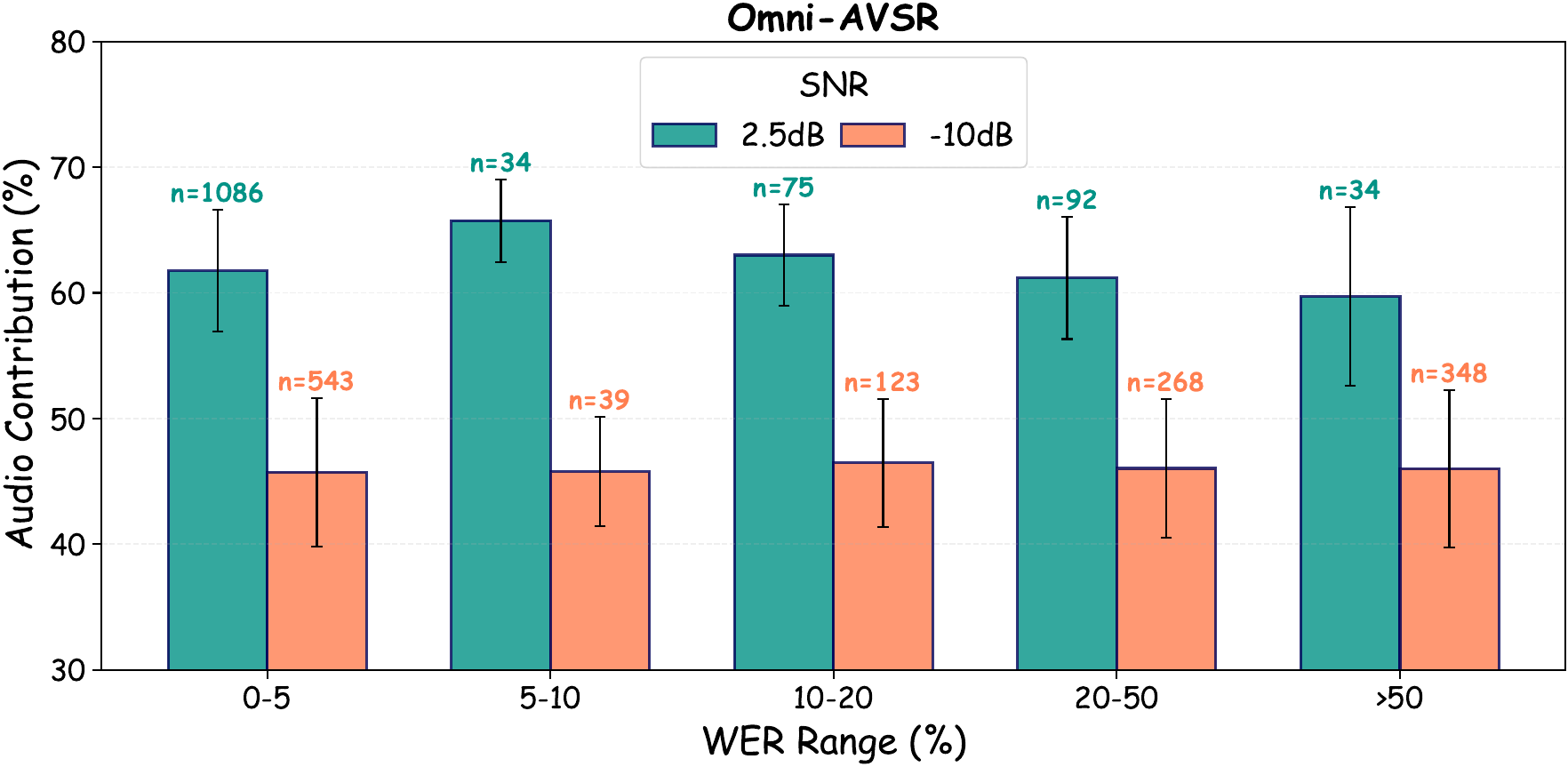}
     \end{subfigure}
     
     \begin{subfigure}[b]{0.35\textwidth}
         \centering
         \includegraphics[width=\textwidth]{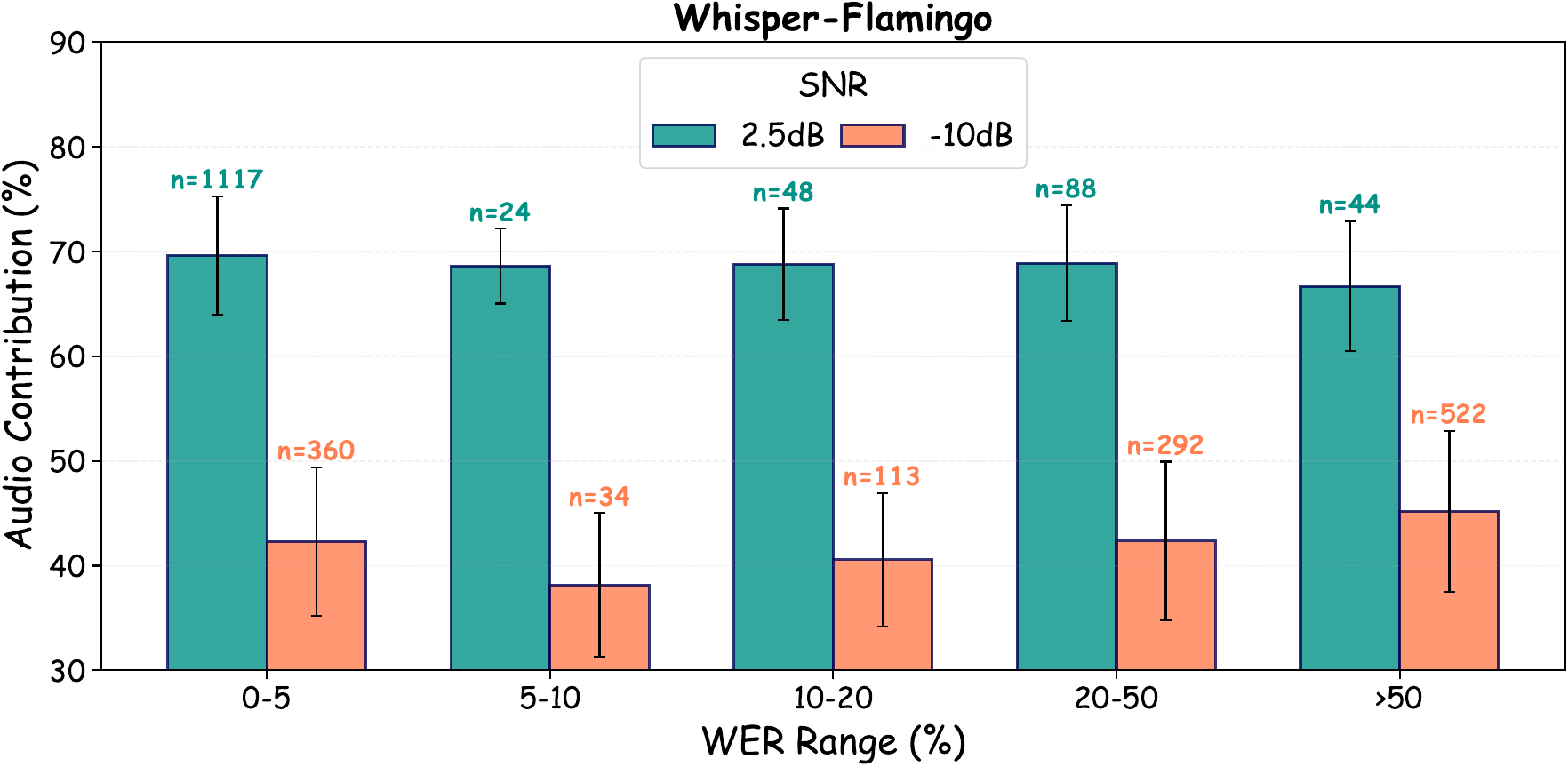}
     \end{subfigure}
    \caption{Mean global audio contribution across WER ranges for Omni-AVSR \textbf{(top)} and Whisper-Flamingo \textbf{(bottom)} under $2.5$~dB and $-10$~dB conditions. Error bars indicate standard deviation; sample counts are annotated above each bar.}
    \label{fig:wer_analysis}
    \vspace{-0.5cm}
\end{figure}

We finally examine whether recognition difficulty influences modality contributions by analyzing audio-visual balance across WER ranges. Figure~\ref{fig:wer_analysis} shows the mean audio contribution across $5$ WER ranges for Omni-AVSR and Whisper-Flamingo under moderate ($2.5$~dB) and severe ($-10$~dB) noise conditions. 

For both models, audio contribution remains stable across WER bins within each SNR condition. Omni-AVSR maintains an audio contribution of around $45$-$46$\% at $-10$~dB and around $60$-$65$\% at $2.5$~dB, regardless of recognition accuracy. Whisper-Flamingo exhibits the same pattern: $38$-$45$\% audio contribution at $-10$~dB and $67$-$70$\% at $2.5$~dB across all WER bins. This suggests that current AVSR models do not adapt their audio-visual weighting based on input difficulty. While we observe some variation in audio contribution across WER bins, these differences remain small (typically within $3$-$7$ percentage points) compared to the shift induced by SNR (up to $27$ points for Whisper-Flamingo). Crucially, no consistent trend emerges linking higher WER to increased or decreased audio reliance. 

\begin{tcolorbox}[enhanced, drop fuzzy shadow, colback=darkmaroon!15, colframe=darkmaroon, title=Finding 6: SNR-Driven Modality Balance, fonttitle=\bfseries] Modality contributions are determined by acoustic conditions rather than recognition difficulty. \end{tcolorbox}

\section{Conclusion}
\label{sec:conclusion}

We presented Dr.\ SHAP-AV, a framework for analyzing modality contributions in AVSR using Shapley values. Through experiments on six state-of-the-art models, we found that: (1) models shift toward visual reliance under noise but maintain surprisingly high audio contributions even under severe degradation; (2) modality balance evolves during token generation, shows architecture-dependent sensitivity to utterance duration, and models preserve temporal alignment between inputs and outputs; and (3) SNR is the dominant factor driving modality balance, while noise type and recognition difficulty induce only minor variations. These findings suggest that explicit mechanisms to modulate modality reliance based on input quality may improve robustness. We also encourage future AVSR works to report modality contributions as a standard diagnostic for understanding multimodal integration.

\section{Generative AI Use Disclosure}
Generative AI tools were used solely for the purpose of checking grammatical errors, spelling correction, and sentence polishing.

\section{Acknowledgements}
We thank Andrew Rouditchenko for his help in integrating Whisper-Flamingo models into Dr.\ SHAP-AV.

\bibliographystyle{IEEEtran}
\bibliography{mybib}

\end{document}